\documentclass[global]{svjour}
\usepackage{graphicx} 
\usepackage{amssymb,amsmath}
\journalname{myjournal}
\begin{document}
\title{Characteristics of Ant-Inspired Traffic Flow}
\subtitle{Applying the social insect metaphor to traffic models}

\titlerunning{Characteristics of Ant-Inspired Traffic Flow}

\author{Alexander John\inst{1} \and Andreas Schadschneider \inst{1,2} 
\and Debashish Chowdhury\inst{3}  \and Katsuhiro Nishinari\inst{4}             
}                     
%
%

\authorrunning{Alexander John et al.}

\institute{Institut f\"ur Theoretische Physik, Universit\"at zu K\"oln, 
50937 K\"oln,  Germany
\and Interdisziplin\"ares Zentrum f\"ur komplexe Systeme, Bonn, Germany
\and Department of Physics, Indian Institute of Technology, 
Kanpur 208016, India
\and Department of Aeronautics and Astronautics, Faculty of Engineering, 
University of Tokyo, Tokyo, 113-8656, Japan
}
\date{Received: date / Revised version: date}
%
\maketitle
\begin{abstract}
  We investigate the organization of traffic flow on preexisting 
  uni- and bidirectional ant trails. Our investigations comprise a
  theoretical as well as an empirical part. We propose 
  minimal models of uni- and bi-directional traffic flow implemented as
  cellular automata. Using these models, the spatio-temporal organization
  of ants on the trail is studied. Based on this, some unusual flow 
  characteristics which differ from those known from other traffic
  systems, like vehicular traffic or pedestrians dynamics, are found.
  The theoretical investigations are supplemented by an empirical
  study of bidirectional traffic on a trail of \textit{Leptogenys
    processionalis}.  Finally, we discuss some plausible implications
  of our observations from the perspective of flow optimization.
\end{abstract}

\section{Introduction}
\label{intro}

Traffic-like problems are relevant in various biological systems ranging
from the motion of motor proteins in cells on a microscopic scale 
\cite{howard,chowd-pattern} to the behavior of social insects 
on a macroscopic scale \cite{burd2,burd3,duss1,duss2}. 
This has led to the development of various models that try to explain
the collective behavior observed empirically in these systems.
Besides their interesting transport properties, colonies of social
insects can be seen as multi-agent systems, facing and solving various
kinds of problems \cite{couzin,duss1,duss2}. For that reason the
social insect metaphor has already been employed with great success in
computer science \cite{intelligence,bonabeau}.  
This can also be extended to other fields, for example models of pedestrians 
dynamics where the concept of chemotaxis \cite{ratnieks,ant-large} 
has been adopted for incorporating the mutual interactions of pedestrians
\cite{bionics}.
A 'virtual chemotaxis' mechanism
allows these models to reproduce the collective phenomena observed
empirically in the dynamics of large crowds. In addition, also effects
arising from the direct coupling of counterflowing streams of
pedestrians or ants have been subject of extensive empirical as well as
theoretical investigations \cite{Encycl08,duss1,duss2,john-ped05}.

Taking into account these basic aspects, which will be addressed in more
detail below, minimal particle-hopping models 
for traffic flow on preexisting uni- and bidirectional ant trails have been
proposed \cite{chowd-uni,john-jtb,john-phd}. These models are
implemented as cellular automata that have been previously used
to successfully describe traffic flow in various other fields, e.g.\
physical systems \cite{chopard} like surface growth and human transport
systems on highways and in cities \cite{nasch,css}.

The main focus of the present work consists in the proposal of
very basic and hence quite general models. We identify the main
features of the flow-density relation of already established trails
and present simulation results for the proposed models in the full parameter
regime of attainable densities.

In addition to the theoretical considerations, we present the first
empirical traffic data from a bidirectional trail of
\textit{Leptogenys processionalis}. In order to facilitate the
comparison with the predictions of the simplified models, the proposed
experimental setup, that is the choice of the ant species and of the type of trail,
has to satisfy certain restrictions. In order to reduce complexity, we 
observed traffic flow of a monomorphic species on a preexisting trail.
More details will be given in Sec.~\ref{emp} and in the subsequent discussion. 

Nevertheless, the techniques employed might also apply
in a more general setup. The data obtained this way are compared with
previous empirical results \cite{burd2,burd3,johnson} and give some
indications for more realistic extensions of the minimal models
studied here which focus only on some key features of ant traffic
(additional aspects related to non-equilibrium physics or the
application to pedestrian dynamics are treated in
\cite{john-tgf05,john-ped05}).

\section{The unidirectional model}

The model for unidirectional flow that we have introduced in
  \cite{chowd-uni} and that will be considered further in this paper
can be seen as an extension of the so-called totally asymmetric simple
exclusion process (TASEP) \cite{schuetz,css}, a well studied model in
non-equilibrium physics. Ants will be modelled by particles
\footnote{In the following we will use the terms ant and particle interchangeably.} that
  move stochastically along a one-dimensional lattice. In the following, we 
define the TASEP and then certain extensions are added that lead to
the unidirectional ant trail model that we propose in this paper.
In a further step the unidirectional model will be extended to the bidirectional one.

\subsection{TASEP regime}

The TASEP is defined by a simple set of rules. Particles 
(representing the ants) hop in a fixed direction from one
lattice-site $i$ to the next one $i+1$.
Each site can be occupied only by one particle (exclusion principle).
Time evolution is continuous, which can be realized by a
random-sequential update scheme: In each update step, one site $i$ is
chosen at random. If site $i$ is occupied and $i+1$ is empty, hopping
takes places with probability $q$.  If site $i+1$ is blocked by
another particle, nothing happens (exclusion principle) and another
site is chosen randomly. A particle leaving the last site $L$ will hop
to site $1$, i.e.\ motion takes places in a ring geometry.  More
realistic geometries are possible, but the main effects in our models
are quite robust against a change of boundary conditions
\cite{kunwar}.

One important way of characterizing a traffic-like system is the
fundamental diagram \cite{may}. It relates the average velocity $V$ or
the flow $F$ with the density $\varrho$ of agents in the system.  Both
relations $F(\varrho)$ and $V(\varrho)$ are equivalent due to the
hydrodynamic relation $F=\varrho V$. Nevertheless, both will be used
in the following since certain features can be seen more clearly using
the average velocity, other using the flow. 

In the TASEP with random-sequential dynamics and $N$ particles
hopping on a lattice with $L$ sites, flow $F$ and average velocity $V$
as a function of density $\varrho$ are exactly given by (see e.g.\
\cite{schuetz,css})
\begin{equation}
V(\varrho)=q(1-\varrho), \qquad F(\varrho)=q\varrho(1-\varrho)
\qquad \text{with}\qquad 
\varrho=\dfrac{N}{L}.
\end{equation}
Mutual blocking of particles is the only mechanism of interaction,
leading to a strictly monotonic decrease of average velocity with
increasing density.  Flow and average velocity are directly linked to
the spatio-temporal distribution of particles. The density $\varrho$
gives the probability of finding a particle at a site $i$, whereas
$1-\varrho$ gives the probability of finding a particular site being
unoccupied. The distribution of particles is homogeneous on
time-average. Due to increased mutual blocking the average velocity
decreases strictly monotonically with increasing density, which agrees
with the behavior found for example in vehicular traffic
\cite{may,css}.

\subsection{Cluster regime}
\label{sec_cluster}

\begin{figure}[ht]
\begin{center}
\includegraphics[width=0.45 \textwidth]{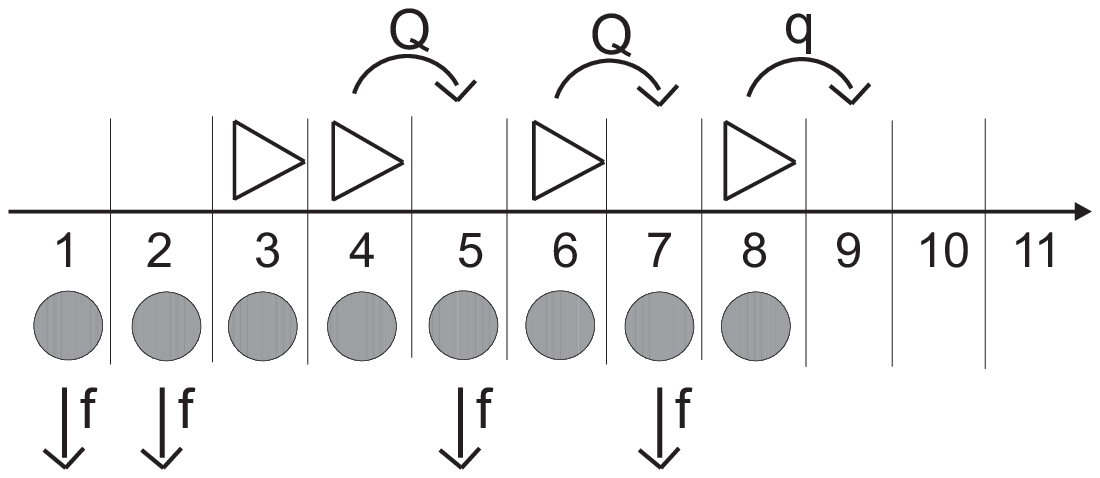}\hfill
\includegraphics[width=0.45 \textwidth]{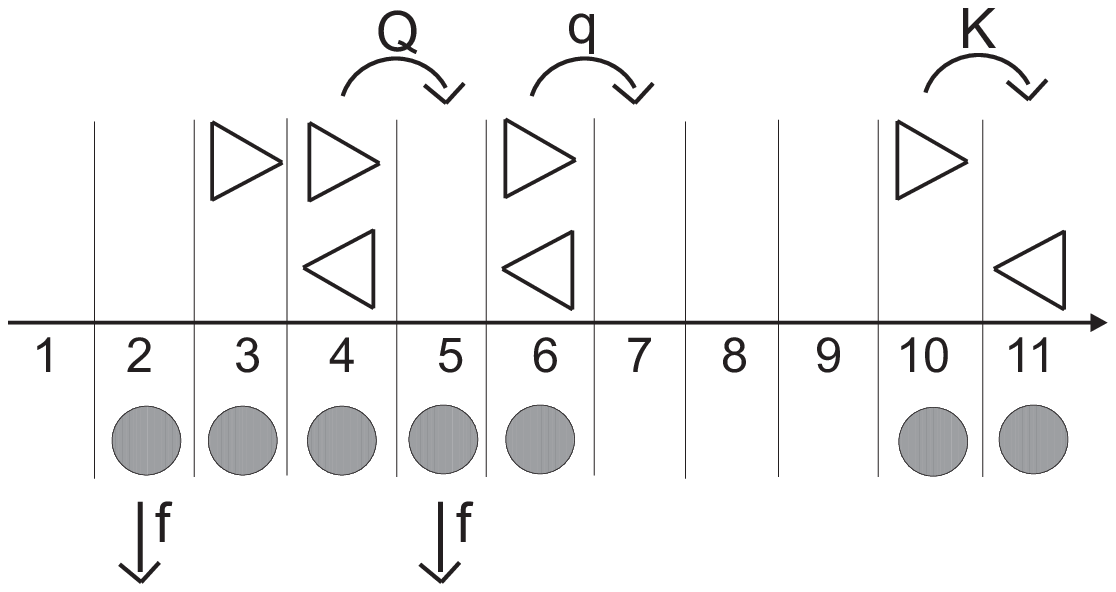}
\end{center}
\caption[]{Definition of the uni- (left) and bidirectional 
  model (right): The symbols correspond to particles (representing
  ants) moving to the right
  $(\triangleright)$, particles moving to the left $(\triangleleft)$, and
  pheromone marks $(\bullet)$. Although a natural trail is
  two-dimensional, the ant motion can be considered to be
  one-dimensional allowing to map the trail onto a one-dimensional
  lattice with $L$ sites. Particles move stochastically along
  the lattice, with hopping rates $q$ and $Q$.  
  Free pheromones can evaporate with rate $f$. Java applets
  illustrating the dynamics of the models can be found at
  \textit{www.thp.uni-koeln.de/$\sim$aj}.}
\label{eps1}
\end{figure}

The model for unidirectional ant trails \cite{chowd-uni} is obtained
by extending the TASEP. Now each lattice site can also be occupied by
a pheromone-mark and/or one ant (see Fig.~\ref{eps1} left).  If a
pheromone mark but no ant is present at site $i+1$, an ant at site $i$
will hop to site $i+1$ with probability $Q$. If neither an ant nor a
mark are present, hopping will take place with probability $q<Q$.
Also pheromones evaporate: if one site is marked but no particle is
present, this mark will be removed (evaporated) with probability $f$.
If a particle is present and the site is chosen for updating, the
mark will not evaporate, reflecting the reinforcement of the
pheromone-mark by the ants. The corresponding probability $p(i,\Delta
t)$ of finding a pheromone at site $i$ is therefore given by
\begin{equation}
p(i,\Delta t)=\left\lbrace   \begin{array}{ll}
1 &\quad \text{if $i$ is occupied}\\ 
(1-f)^{\Delta t}& \quad\text{else}\\ 
\end{array}      \right.  
\label{evap}
\end{equation}
where $\Delta t$ is the time-interval between the occupation of a site
$i$ by two succeeding ants, called time-headway in traffic
engineering. Obviously the time-headway determines the decay of the
probability of finding a pheromone.

\begin{figure}[ht]
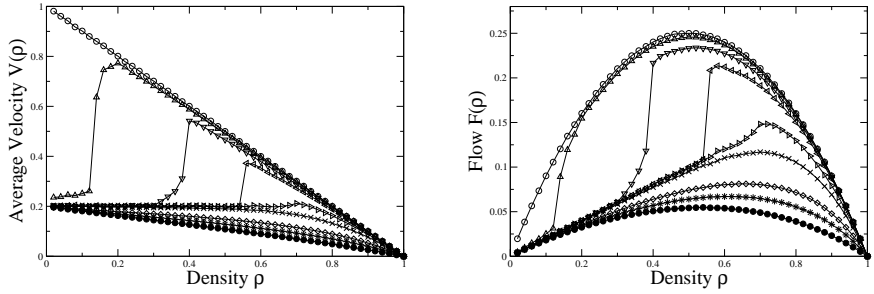

\begin{center}
\includegraphics[scale=0.22]{uni-vel.eps}\hfill
\includegraphics[scale=0.22]{uni-flow.eps}
\end{center}
\caption[]{Fundamental diagrams for the cellular automaton implementation 
  of the unidirectional model: $Q=0.9$, $q=0.2$ and $f=0(\circ$),
  0.0002($\vartriangle$), 0.0008($\triangledown$),
  0.002($\triangleleft$), 0.008($\triangleright$), 0.02 ($\times$),
  0.08($\diamond$), 0.2($\ast$), 1($\bullet$). In case of $f=1$ and
  $f=0$ the average velocity $V(\varrho)$ and flow $F(\varrho)$ are
  exactly known from the TASEP.  At low to intermediate densities, the
  average velocity stays constant for a suitable choice of evaporation
  probability $f$.}
\label{eps2}
\end{figure}

In the case $f=0$ $(f=1)$ pheromones will never (immediately)
evaporate and the TASEP-case with hopping probability $Q$ ($q$)
is recovered (Fig.~\ref{eps2}). More interesting properties
arise in case of $0<f<1$. The most surprising feature is the
non-monotonic dependence of the average velocity on density for small
evaporation rates $f$ (Fig.~\ref{eps2} left).

\begin{figure}[ht]
\begin{center}
\includegraphics[scale=0.14]{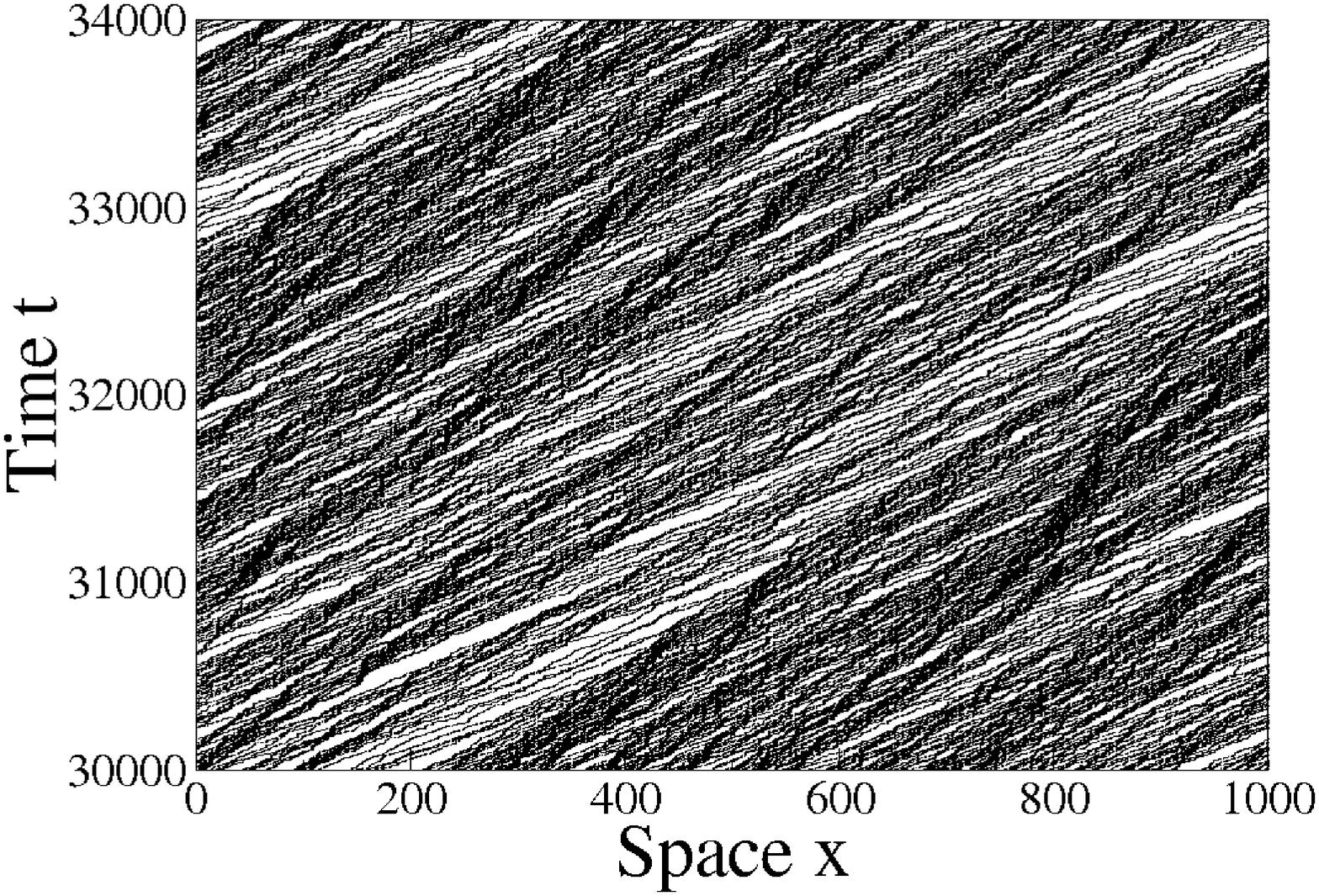}\hfill
\includegraphics[scale=0.14]{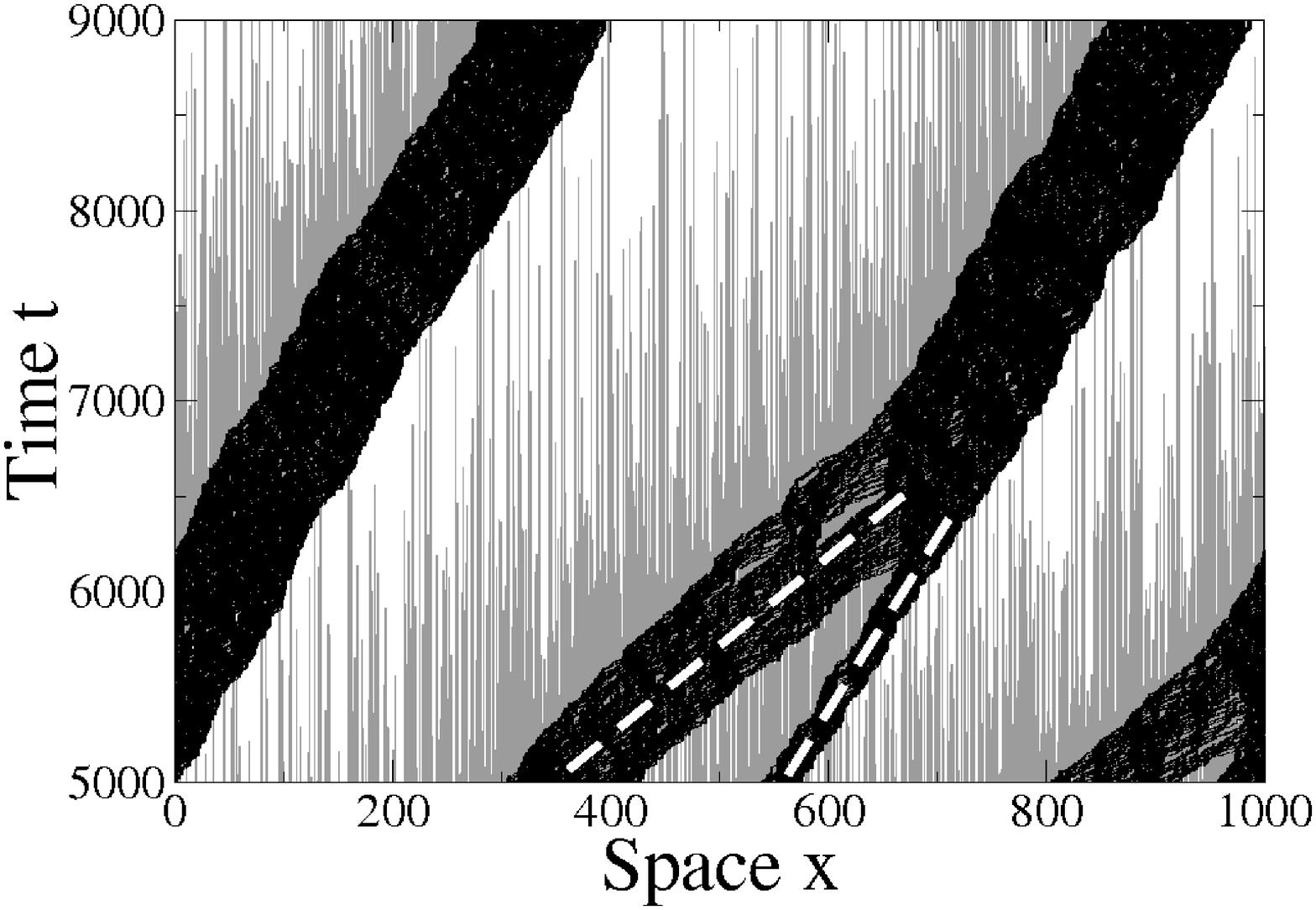}\hfill
\includegraphics[scale=0.14]{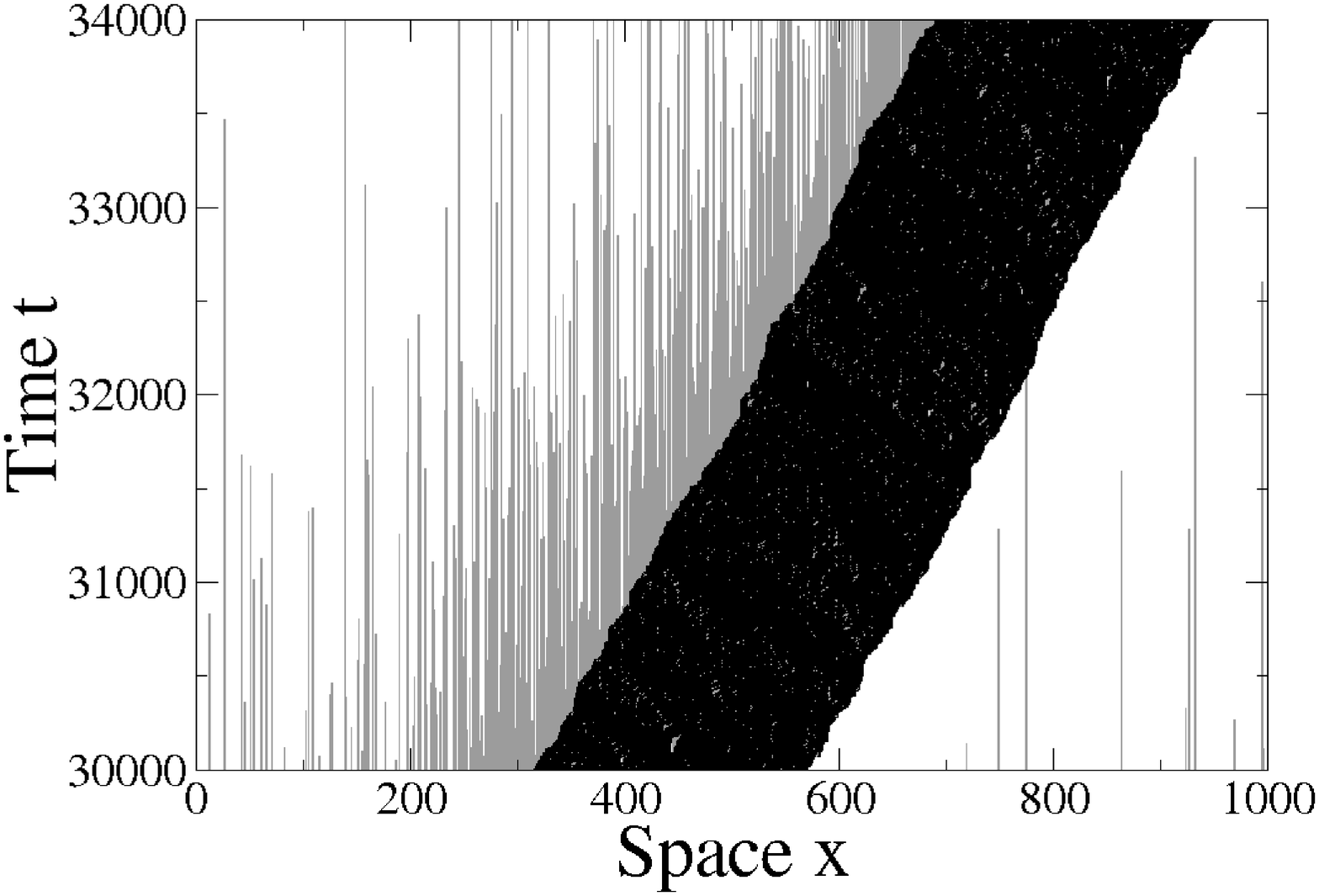}

\end{center}
\caption[]{Space-time plots for the unidirectional model: $Q=0.9$, $q=0.2$, 
  $f=0.001$, $\varrho=0.2$. On the left the TASEP case $(p=0.9)$ is
  shown. Besides fluctuations, particles are distributed randomly. The
  other two plots show different stages of the cluster formation in
  the unidirectional ant trail model. Each ant (black) is followed by a
  trace of pheromones (grey). Due to evaporation this trace has a
  finite length. As indicated by the different slopes of the ants
  trajectories (middle) clusters move with different velocities
  depending on the presence of pheromones. Finally, in the stationary state
  only one moving cluster survives (right). 
}
\label{eps3}
\end{figure}

At low to intermediate densities, the average velocity stays constant
($V=q$).  Beyond a certain threshold value a sharp increase of
velocity can be seen. In the regime of high densities, the monotonic
decrease known from the TASEP is found. Both regimes are a consequence
of the incorporation of pheromone marks. Each ant is followed by
a trace of marks (Fig.~\ref{eps3} middle). A succeeding ant
perceiving this trace will hop with probability $Q>q$. If the
preceding particle sees no pheromone mark, it will hop with probability
$q<Q$. Then faster ant will catch up with the slower ones
forming a moving cluster (see Fig.~\ref{eps3} middle and
Fig.~\ref{eps3} right). Similar results are known from systems with
particle-wise defects \cite{ferrari} where each particle $j$ has its
own individual hopping probability $q_{j}$. Here also clusters
are formed and the average velocity stays constant $v=q$. An
example from vehicular traffic is a platoon of cars following a
slow truck. At very low evaporation rates or at high densities (small
average distance or distance-headway), all ants perceive pheromone
marks. So on average their hopping probabilities all reach the same
value $Q$, leading to TASEP-like features. The corresponding
distribution of ants becomes homogeneous (see Fig.~\ref{eps3} left).

\subsection{A comparison with models of vehicular traffic}
\label{vergleich}

As already mentioned, cellular automaton models are well established
in case of vehicular traffic \cite{css}. They seek to capture the complex
behavior of a system composed of humans driving in their cars. Besides
certain similarities, models of ant traffic should also reflect
crucial differences to vehicular traffic. One is the lack of
some kind of velocity memory. In our model, ants reach their walking
speed within one single update step. This is similar to the motion of
pedestrians where the walking speed is also reached within a time less
than 1 sec. In contrast, in vehicular traffic cars cannot accelerate
instantaneously to the maximum velocity and so velocity is only
increased gradually \cite{nasch,bionics,css}.

Furthermore, the behavior of ants can be expected to be much more
homogeneous than that of drivers on a highway. Here a mixture of
different vehicles with different maximum speeds, acceleration
capabilities, and so on, but also of different kinds of drivers with
different moods or attitudes, influences the overall behavior
\cite{may}. Therefore, the range of possible velocities in vehicular
traffic can be assumed to be larger than in ant traffic, at least in
comparison with monomorphic species. As a consequence of the speed
homogeneity, overtaking is rarely observed on ant trails at least
under certain conditions \cite{john-phd} and has therefore not been
included into our model.

Also the definition of the update procedure itself can have a strong
influence on the dynamics. For modelling vehicular traffic a
time-parallel update is widely used which performs a synchronous
update of all lattice sites \cite{css}. As only the occupation of the
lattice at the moment of updating is used, a synchronous update
incorporates time latencies of the drivers. But in case of ants the
perception range is limited to their immediate environment
\cite{ant-large,moffet,ratnieks} and the random-sequential update
described above is more appropriate.

The stochasticity in the model has mainly two sources.  First, many
influencing factors are not known or difficult to include explicitly
since the model would become too complicated.  Thus they are included
in a statistical sense through probabilities for a certain behavior.
The second reason lies in the behavior of the ants themselves.  They
appear to possess an intrinsic stochasticity depending on the
evolution of the particular species \cite{ant-large}. This is also
widely used in applications \cite{intelligence,bonabeau}. In
vehicular traffic stochasticity is used to incorporate fluctuations in
the driver's behavior which can lead to spontaneously formed phantom
jams \cite{nasch}.


\section{The bidirectional model}

For extending the unidirectional model to the case of bidirectional
flow we have investigated several models \cite{john-ped05,john-jtb}.
As one common requirement they should reduce to the unidirectional
model in case of vanishing counterflow.  In the extension discussed
here another lattice for ants moving in the opposite direction is
added (see Fig.~\ref{eps1} right). Both lattices for ants obey the
same set of dynamical rules. An extension to multiple lanes is then
straightforward by adding further one-dimensional lattices.
Experimental studies already addressed this issue \cite{couzin,burd3}.
Going further one could consider the extension to a two-dimensional
model. Here the trail topology including the formation of lanes itself
could be incorporated more naturally. Especially the relation between
flow properties and trail topology could be investigated. Also the
analogies to already existing models of pedestrians dynamics surely
provide a fruitful perspective \cite{bionics,Encycl08}.

Overall, one deals with three hopping probabilities depending on the
occupation of the nearest neighboring site in hopping direction. In
absence of counterflow again the rates $q$ and $Q$ apply. In case of
counterflow one additional rate $K$ (see Fig.~\ref{eps1} right) is
used. This is a crucial difference to most models of vehicular traffic
which usually neglect the coupling of lanes in opposing directions
\cite{css}. Workers facing each other in opposite directions have to
slow down due to information exchange or just in order to avoid
collision \cite{burd2,burd3,couzin,duss2}.  Therefore, we choose
hopping probabilities satisfying $K<q<Q$. 

A related problem is the organization of ant traffic at bottlenecks.
For \emph{Lasius niger} this has already been investigated
experimentally \cite{duss2}.  In that case traffic flow is organized
such that the number of encounters inside the bottleneck is reduced.
This is quite different from our bidirectional model where the main
effect originates from these head-on encounters (see Fig.~\ref{eps7}
right). Here some similarities to models of pedestrians dynamics can be drawn
\cite{Encycl08,john-ped05}.  Under crowded conditions separated lanes
for each direction are formed dynamically. These lanes are stabilized
by incorporating the desire of pedestrians to reduce the number of
encounters with others in the counterflow.

\subsection{Equal densities}

In general, the fundamental diagram can be expected to depend on
both densities $\rho_{L}$ and $\rho_{R}$ of left- and right-moving
ants. That is, it has the form $F=F(\rho_{L},\rho_{R})$. For
simplicity, we start the discussion with equal densities in both
directions. Finally, we investigate the full fundamental diagram
focusing on the generic properties and differences to the other cases.

\begin{figure}[ht]
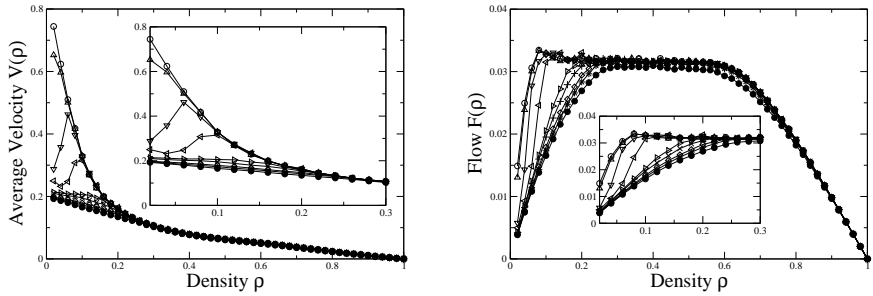

\begin{center}
\includegraphics[scale=0.22]{bi-vel.eps}\hfill
\includegraphics[scale=0.22]{bi-flow.eps}
\end{center}
\caption[]{Fundamental diagrams for the cellular automaton 
  implementation of the bidirectional model: $Q=0.9$, $q=0.2$, $K=0.1$ and
  $f=0(\circ$), 0.0002($\vartriangle$), 0.0008($\triangledown$),
  0.002($\triangleleft$), 0.008($\triangleright$), 0.02 ($\times$),
  0.08($\diamond$), 0.2($\ast$), 1($\bullet$). The average velocity
  shows the same non-monotonicity already observed in the
  unidirectional model. The main property of the bidirectional model
  is exhibited by flow.  For intermediate densities the flow is nearly
  independent of the density $\varrho$ and the evaporation rate $f$.}
\label{eps4}
\end{figure} 

\subsubsection{Cluster regime}

\begin{figure}[t]
\begin{center}
\includegraphics[scale=0.14]{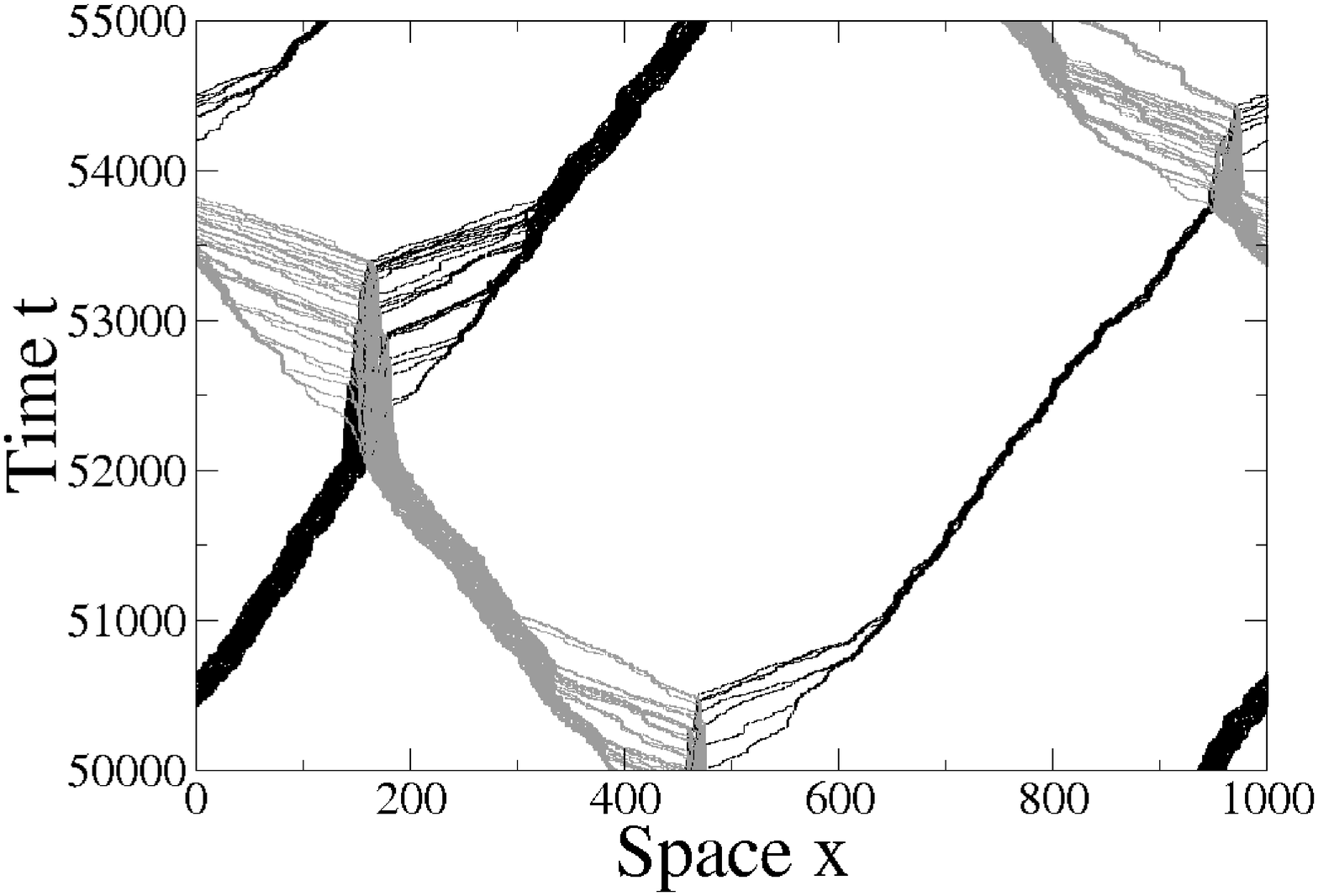}\hfill
\includegraphics[scale=0.14]{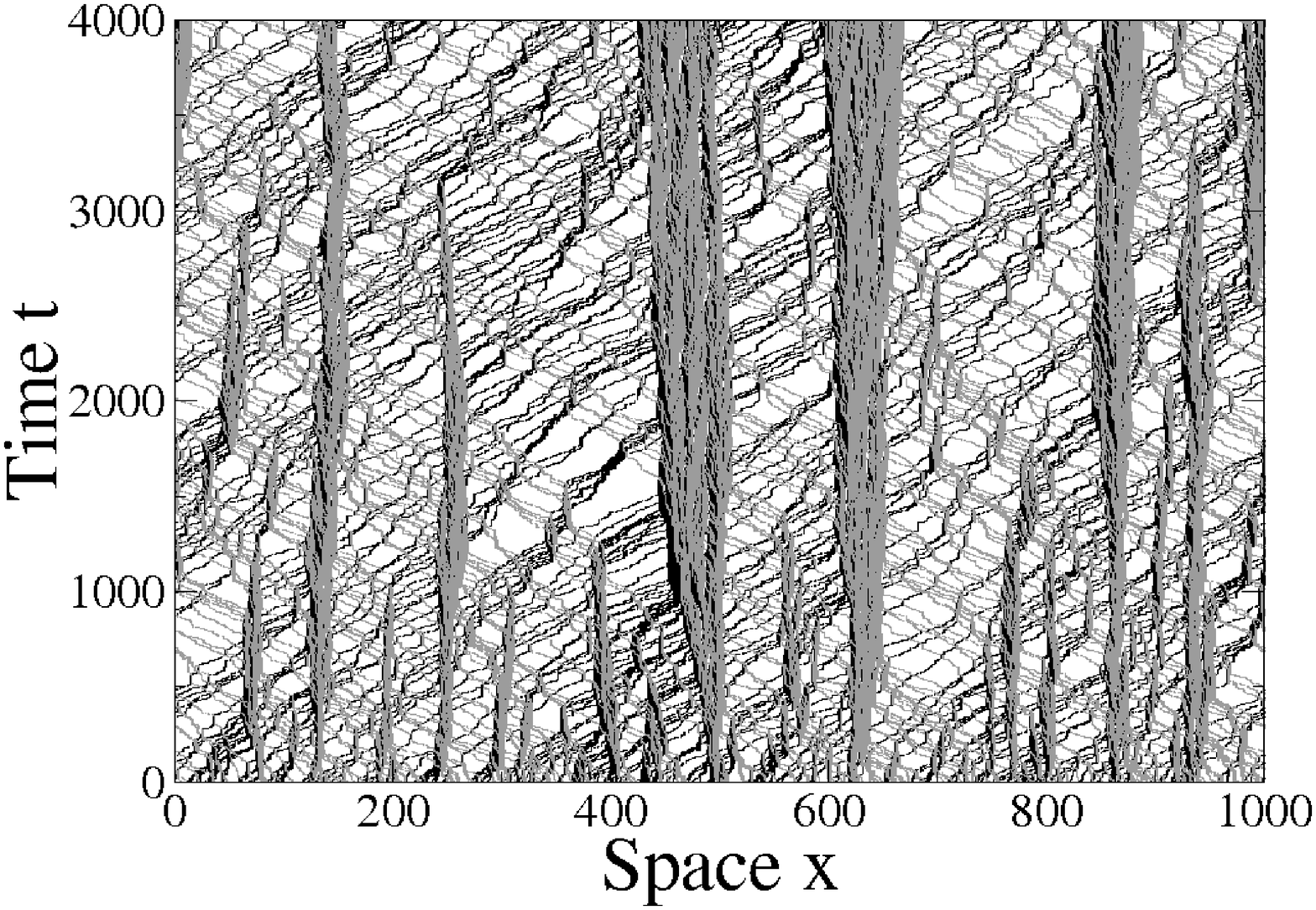}\hfill
\includegraphics[scale=0.14]{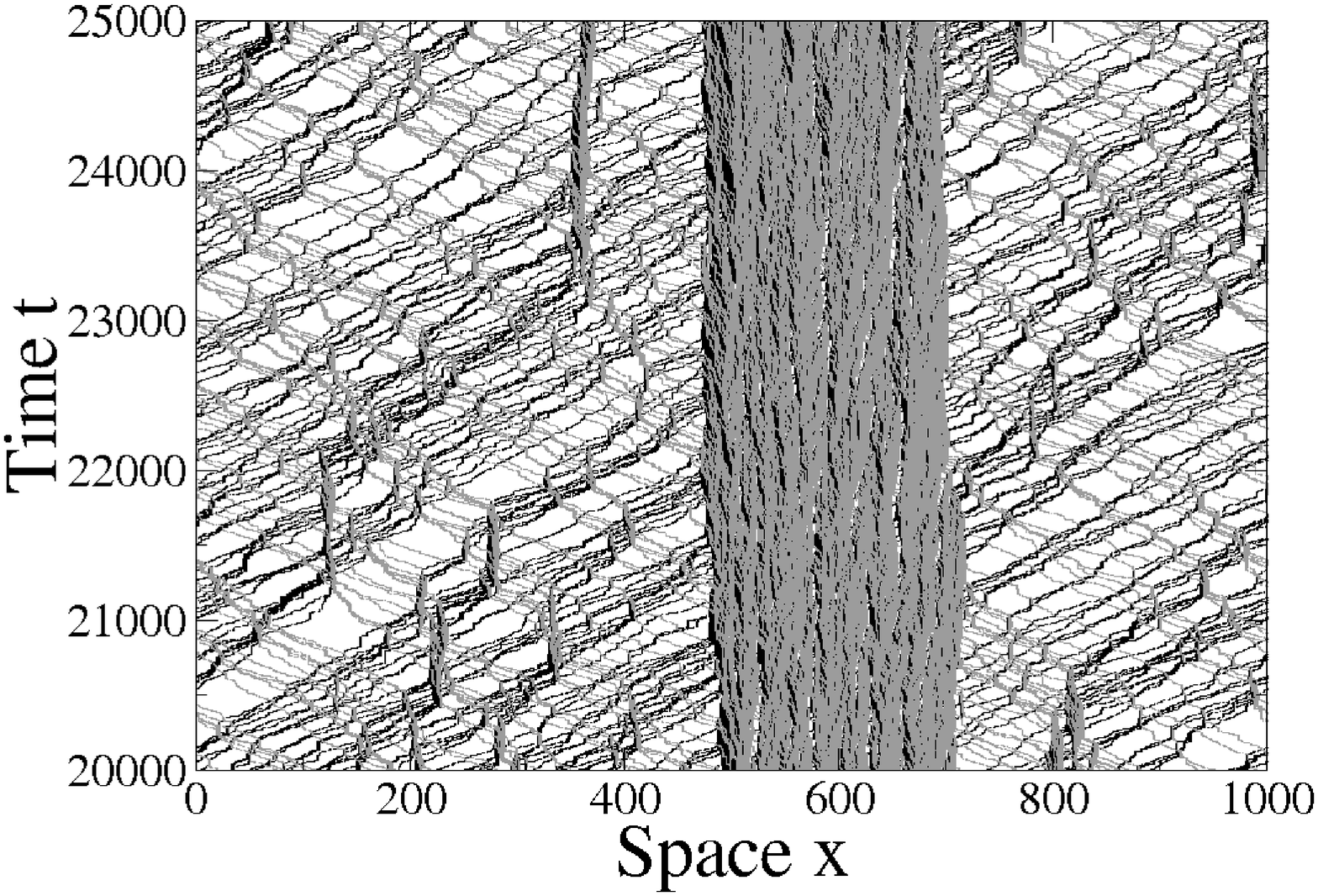}
\end{center}
\caption[]{Space-time plots  for the cellular automaton 
  implementation of the bidirectional model: $Q=0.9$, $q=0.2$,
  $K=0.1$, $f=0.002$. The left plot shows the formation of small
  moving clusters for $\varrho_{L}=\varrho_{R}=0.03$. One observes one
  left-moving (grey) and two right-moving clusters (black). With
  increasing density this regime vanishes as the pheromones are
  present at any site.  Due to the coupling to counterflow a large
  localized cluster emerges (plotted for $\varrho_{L}=\varrho_{R}=0.2$
  in the middle and on the right).}
\label{eps5}
\end{figure}

The average velocity roughly shows the behavior already known from the
unidirectional case (see Fig.~\ref{eps2} left), including the
anomalous density dependence (see Fig.~\ref{eps4} left). As the
average lifetime of the pheromone marks is determined by the mean
distance-headway (mean ant-ant distance) of ants in both directions,
this regime only exists at very low densities.  Similarly to the strictly
unidirectional case, moving clusters for each direction are formed
(see Fig.~\ref{eps5} left). With increasing density the lifetime of
the pheromone marks increases such that they become present at every
site and thus suppressing the mechanism of cluster formation.

\subsubsection{Plateau regime}

The generic property of this model is found in flow and cannot be
observed directly in the average velocity \cite{john-jtb,john-phd}.
For all values of $f$, flow roughly stays constant over a certain
density regime (see Fig.~\ref{eps4} right). Due to the behavior of the
average velocity one observes a shift of the beginning of the plateau
in flow to lower densities with decreasing $f$. But at intermediate to
high densities, there is obviously no dependence on $f$.  This effect
originates from the mutual hindrance by counterflowing ants. Similar
effects are known from systems with lattice-wise disorder \cite{trip}
where the hopping probabilities depend on the actual position (e.g.,
due to construction works or car accidents in vehicular traffic). In
our case we find several localized clusters of ants in both directions
(see Fig.~\ref{eps5} right). They form some kind of dynamically
induced defects, as ants facing these clusters move with reduced
hopping probability $K$.

\subsection{Different densities}

For natural trails, different densities $\rho_{R}$ and $\rho_{L}$ for
each direction are more realistic. Ants moving back to the nest might
be carrying load and therefore will behave quite differently from
outbound ants \cite{burd2,couzin,duss3}.  More complex models could
incorporate this by also employing different hopping rates depending
on the direction. But here we consider only symmetric hopping rates
and focus on right-moving ants, treating the left-moving ones as
counterflow. This is no restriction due to the symmetry
\begin{equation}
F^{R}(\rho_{L},\rho_{R})=F^{L}(\rho_{R},\rho_{L})
\label{f-sym}
\end{equation}

\begin{figure}[ht]
\begin{center}
\includegraphics[scale=0.15]{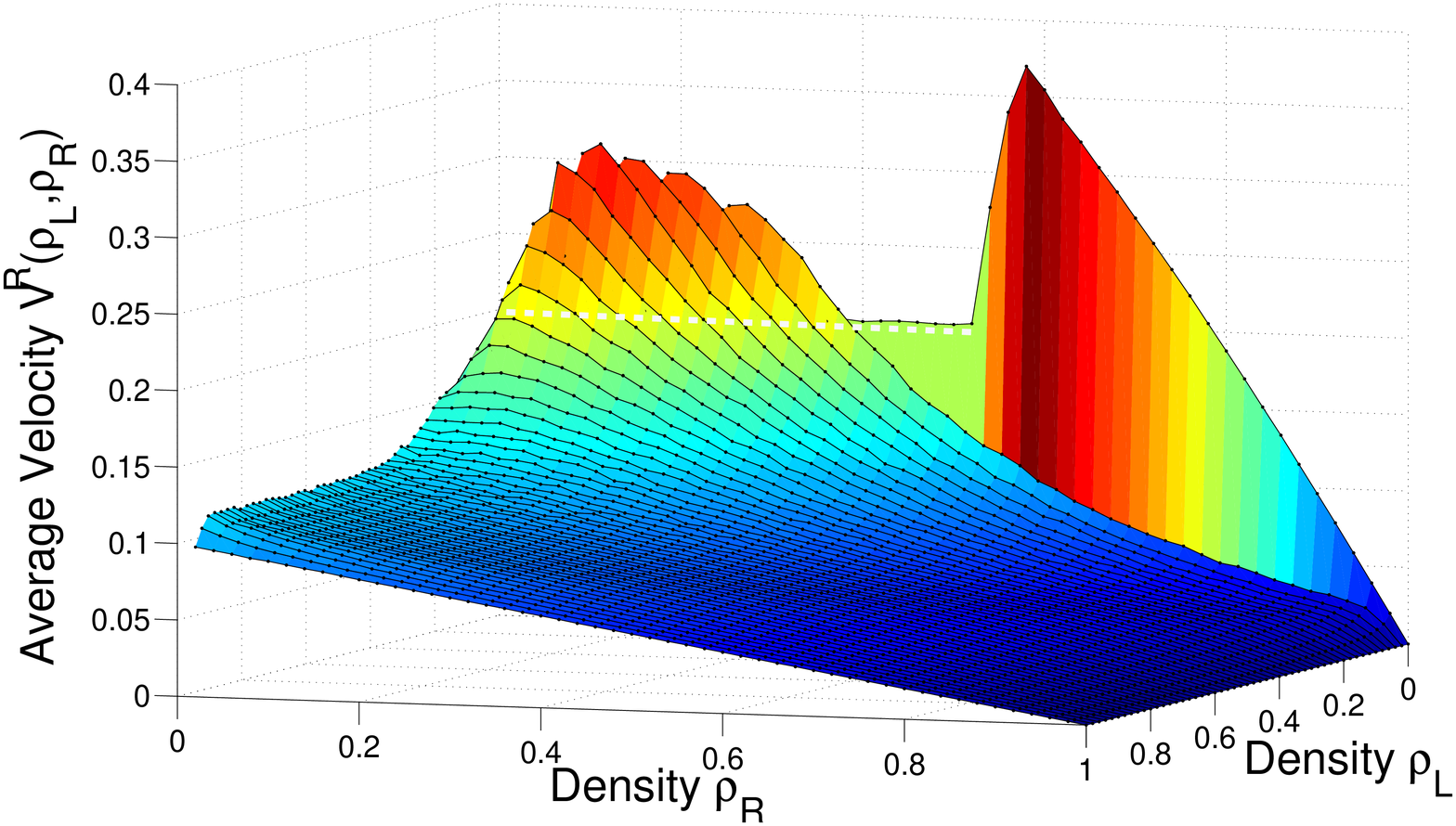}\hglue20pt
\includegraphics[scale=0.15]{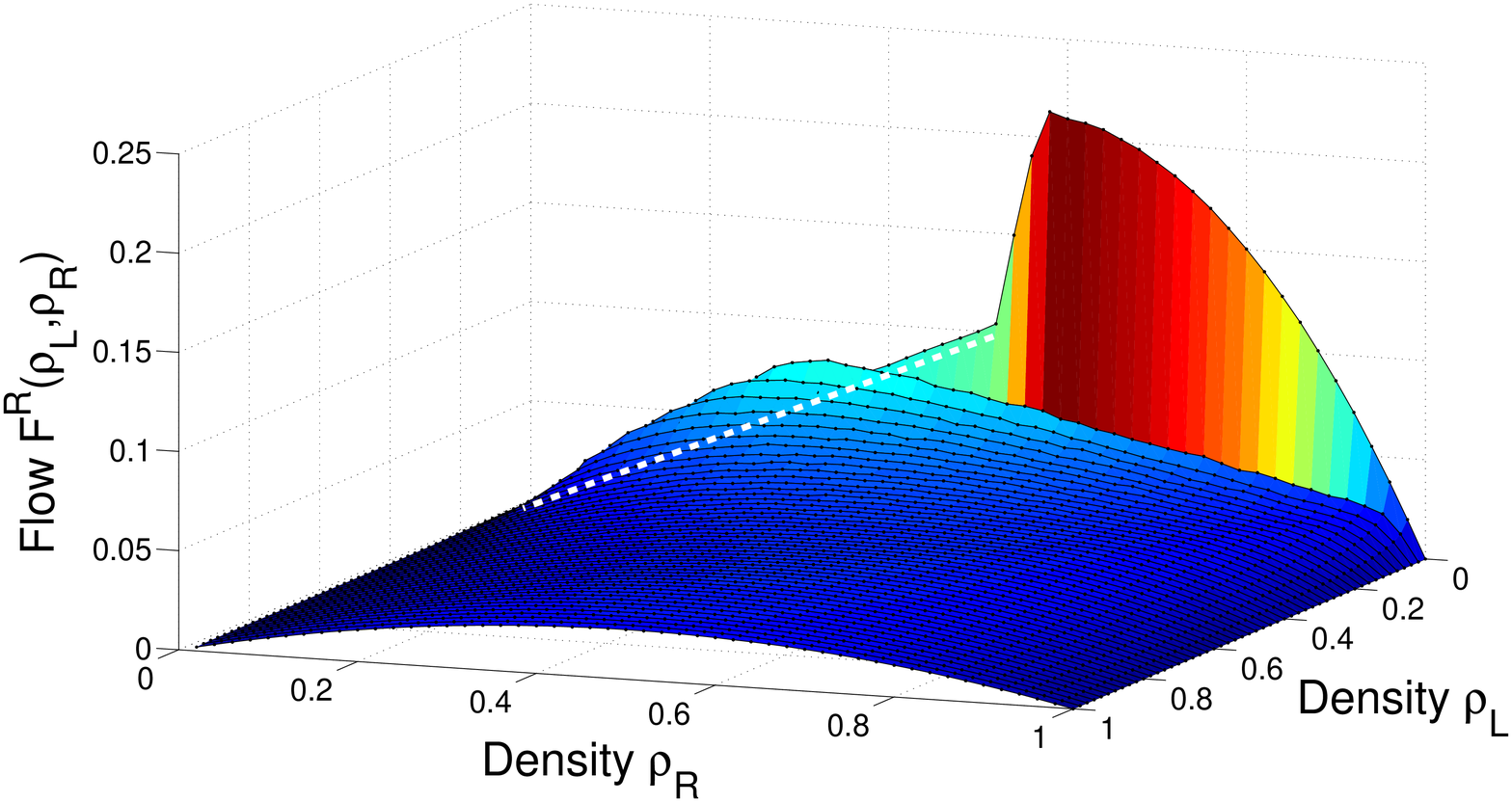}
\end{center}
\caption[]{Fundamental diagrams for the full bidirectional model. 
  Average velocity $V^{R}(\rho_{L},\rho_{R})$ and flow
  $F^{R}(\rho_{L},\rho_{R})$ are shown for the full range of densities
  $(\rho_{L},\rho_{R})\in[0,1]\times [0,1]$. In case of vanishing
  counterflow ($\rho_{L}=0$) obviously the unidirectional case is
  recovered.}
\label{eps6a}
\end{figure}

\begin{figure}[ht]
\begin{center}
\includegraphics[scale=0.15]{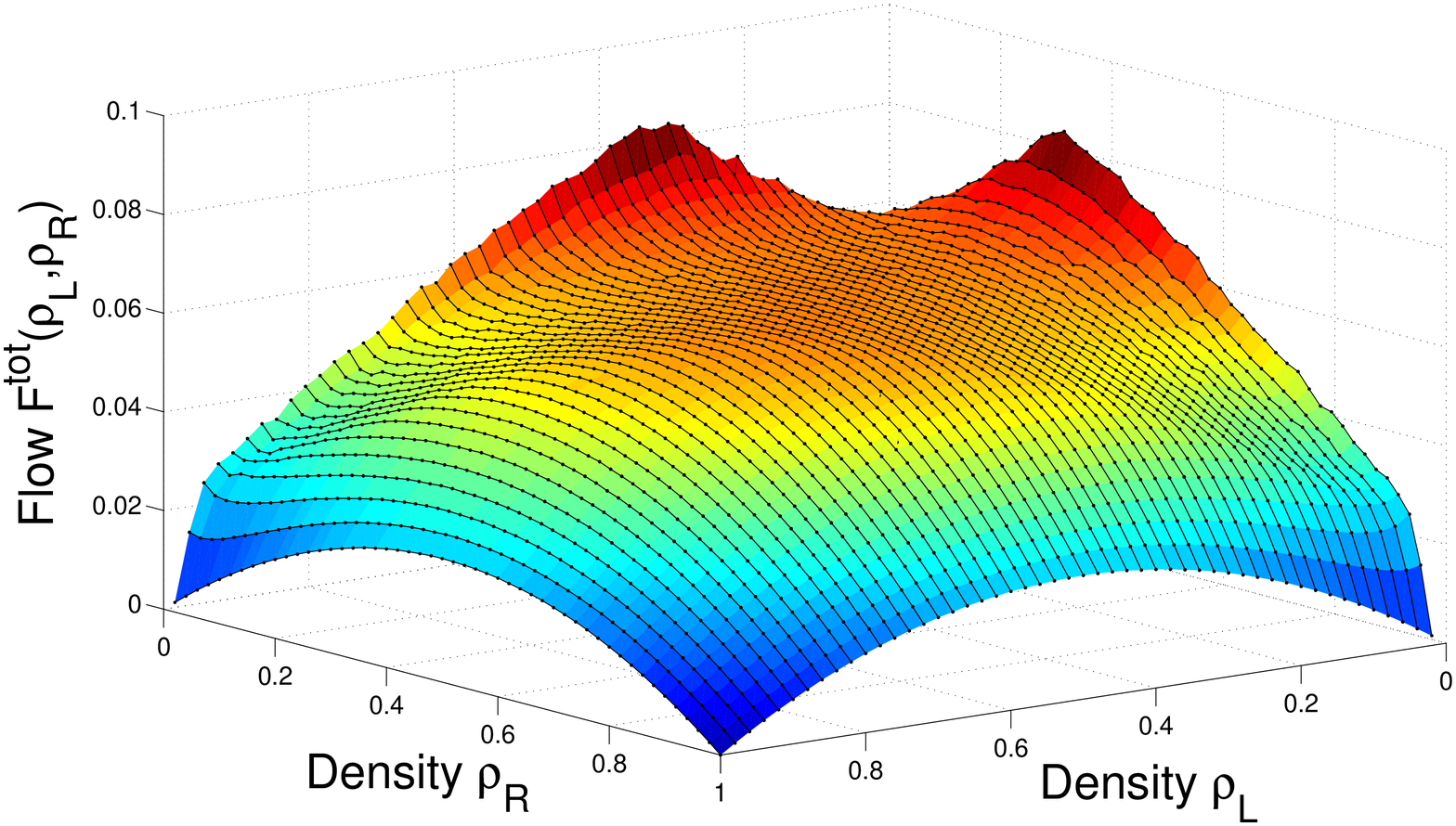}\hglue20pt
\includegraphics[scale=0.15]{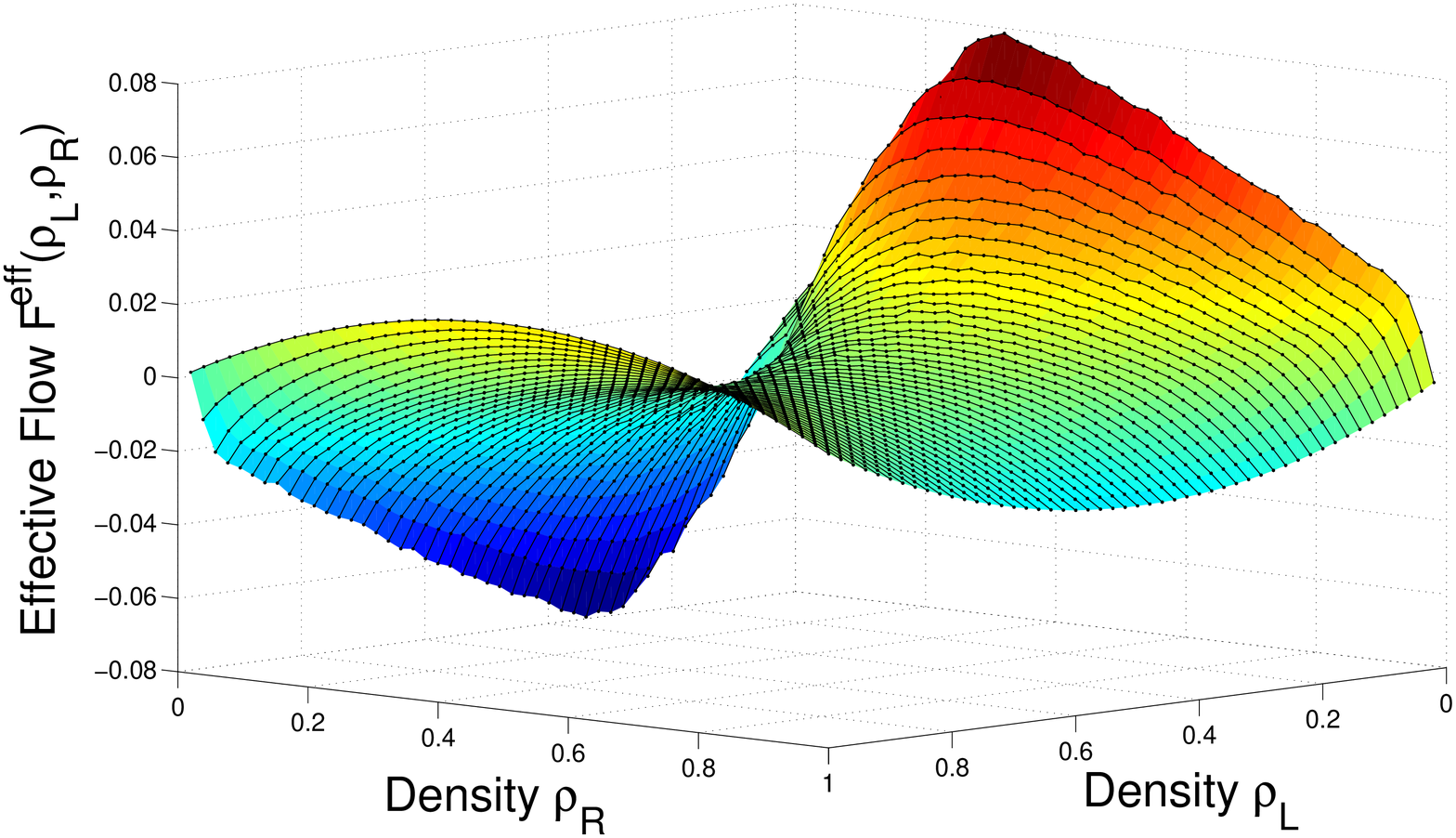}
\end{center}
\caption[]{Extended fundamental diagrams for the full bidirectional 
  model. Total flow $F^{\rm tot}(\rho_{L},\rho_{R})$ and effective flow
  $F^{\rm eff}(\rho_{L},\rho_{R})$ are shown for the full range of
  densities $(\rho_{L},\rho_{R})\in[0,1]\times [0,1]$. Symmetries
  according to (\ref{ftot}) and (\ref{feff}) are found. For clarifying
  the generic features due to counterflow, the limiting cases
  $\rho_{R}=0$ and $\rho_{L}=0$ are not shown.}
\label{eps6b}
\end{figure}

\subsubsection{Properties of the full fundamental diagram}

The behaviour of the average velocity shows strong deviations from the
characteristics of strictly unidirectional flow (see Fig.~\ref{eps6a}
left). Obviously, the regime of constant velocity only exists for a
small area in the $\rho_{L},\rho_{R}$-plane. Although counterflow
leads to a reduced hopping rate $K$, the average velocity is even
increased by counterflow. As already observed for the special case of
equal particle numbers, a non-monotonicity is found at very low
densities $(\rho_{L},\rho_{R})\in]0,0.1]\times]0,0.1]$. For larger
densities, a strictly monotonic decrease is found. As a result, the
average velocity in case of counterflow is lower than for strictly
unidirectional flow ($\rho_{L}=0$ or $\rho_{R}=0$). Overall, the
observed feature is rather an effect of the shared pheromone lattice
than of the reduced hopping rate due to counterflow. As pheromones
become present at nearly any site, no particle-wise disorder can be
induced anymore. Therefore a quasi-TASEP case is attained in which the
impact of counterflow can be neglected. With increasing density of
counterflowing ants the impact of pheromones vanishes and hopping
predominantly takes place with rate $K$. As already observed in the
special case $\rho_{L}=\rho_{R}$, flow reaches a nearly constant
value (see Fig.~\ref{eps6a} right).

\subsubsection{Additional quantities of interest}

Total and effective flow are non-zero quantities in case of different
particle numbers or hopping rates for both directions. The total flow
$F^{\rm tot}(\rho_{L},\rho_{R})$ measures the total number of 
moving ants per unit time and is independent of the direction of
movement.  For quantifying the effective number of ants moving in one
particular direction the corresponding flow $F^{\rm
  eff}(\rho_{L},\rho_{R})$ is used:
\begin{eqnarray}
  F^{\rm tot}(\rho_{L},\rho_{R})  = & 
  F^{R}(\rho_{L},\rho_{R})+  F^{L}(\rho_{L},\rho_{R})
  = F^{\rm tot}(\rho_{R},\rho_{L})\,,
  \label{ftot} \\ 
  F^{\rm eff}(\rho_{L},\rho_{R})  = & F^{R}(\rho_{L},\rho_{R}) -
  F^{L}(\rho_{L},\rho_{R}) = -F^{\rm eff}(\rho_{R},\rho_{L})\,.
  \label{feff}
\end{eqnarray}
As a consequence of employing the same set of hopping rates for both
directions, the quantities exhibit different kinds of symmetries due
to (\ref{f-sym}). One also observes plateaus which are a consequence
of the plateaus found in $F^{R}$ and $F^{L}$ for an intermediate range
of densities $(\rho_{L},\rho_{R})$ (see Fig.~\ref{eps6b}). Both
features might be subject of first qualitative empirical
investigations. The total flow might be of interest especially in the
ecological context of collective transport like foraging \cite{burd2,burd3}
since it measures the total performance of the employed foraging strategy, in contrast
to the effective flow. On the other hand, the latter allows to identify
the overall direction of movement.


\section{A brief outlook to empiricism}
\label{emp}

The discussed analogies between ant- and vehicular traffic can also be
used for experimental investigations. Based on this we collected
qualitative and quantitative traffic data \cite{john-phd}. As
our models are quite simple, the experimental setup, as well as the
observed species, have to meet some well defined requirements. The
observed trail section has to exhibit a constant shape for the period
of data collection which is the equivalent of a static
road. To minimize complexity, intersections or branchings are
excluded. These would be analogous to on- and off-ramps which are known
to have a strong influence on vehicular traffic \cite{css,may}. 
As already pointed out (see Sec.~\ref{vergleich}) the moving agents in
both systems can be assumed to be quite different. For our investigations
we chose ants of the monomorphic species \textit{Leptogenys processionalis}. 
The analogy in vehicular traffic would be a traffic flow composed out of many 
copies of the same car driven by nearly the same driver. This choice ensures that all
moving agents are basically equivalent at least for each direction of
movement. Especially no laden ants were observed.

\subsection{Methods}

The field studies were performed on the campus of the \textit{Indian
  Institute of Science, Bangalore, India}. We obtained qualitative as
well as quantitative data for one bidirectional single lane trail
which was observed for about 25 minutes.  Similar investigations
have been made for ten other trails of the same type which have been
found to exhibit the same qualitative and  quantitative properties.
The criteria employed for the choice of the trails are for
example the trail topology (single-lane, bidirectional) or the observed
means of interaction (e.g., head-on encounters). Particular features
will be addressed in more detail in the subsequent discussion.
\begin{figure}[]
\begin{center}
\includegraphics[scale=0.9]{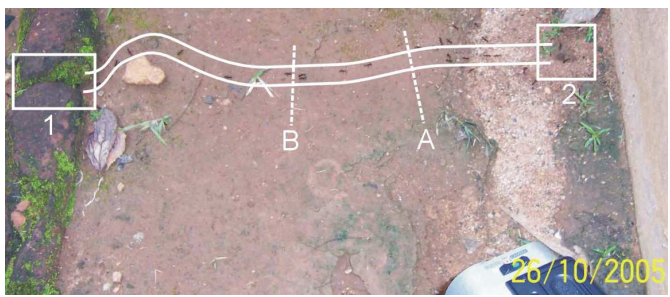}\hfill
\end{center}
\caption[]{Illustration of the experimental setup. The 
  observed section A-B used for data extraction is located in the
  middle of the bidirectional trail. A narrow passage on the left leads to a huge
  lawn (framebox 1). On the right the entrance to the nesting site
  can be seen (framebox 2). 
}
\label{eps7b}
\end{figure}

\begin{figure}[]
\begin{center}
\includegraphics[scale=0.11]{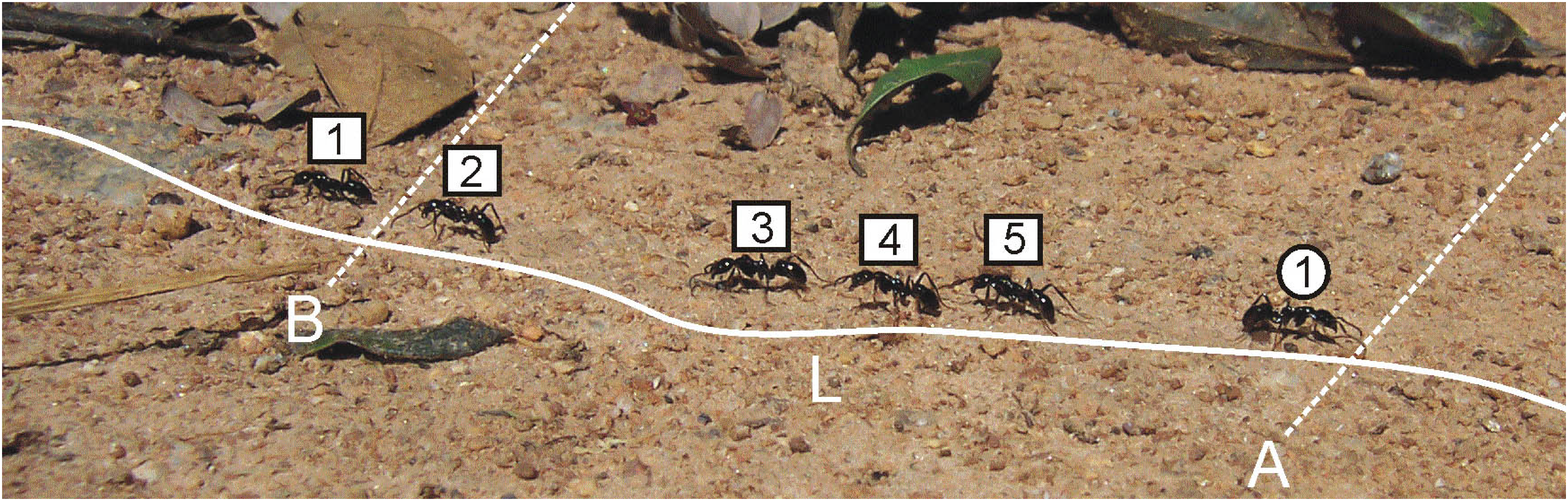}\hfill
\includegraphics[scale=0.145]{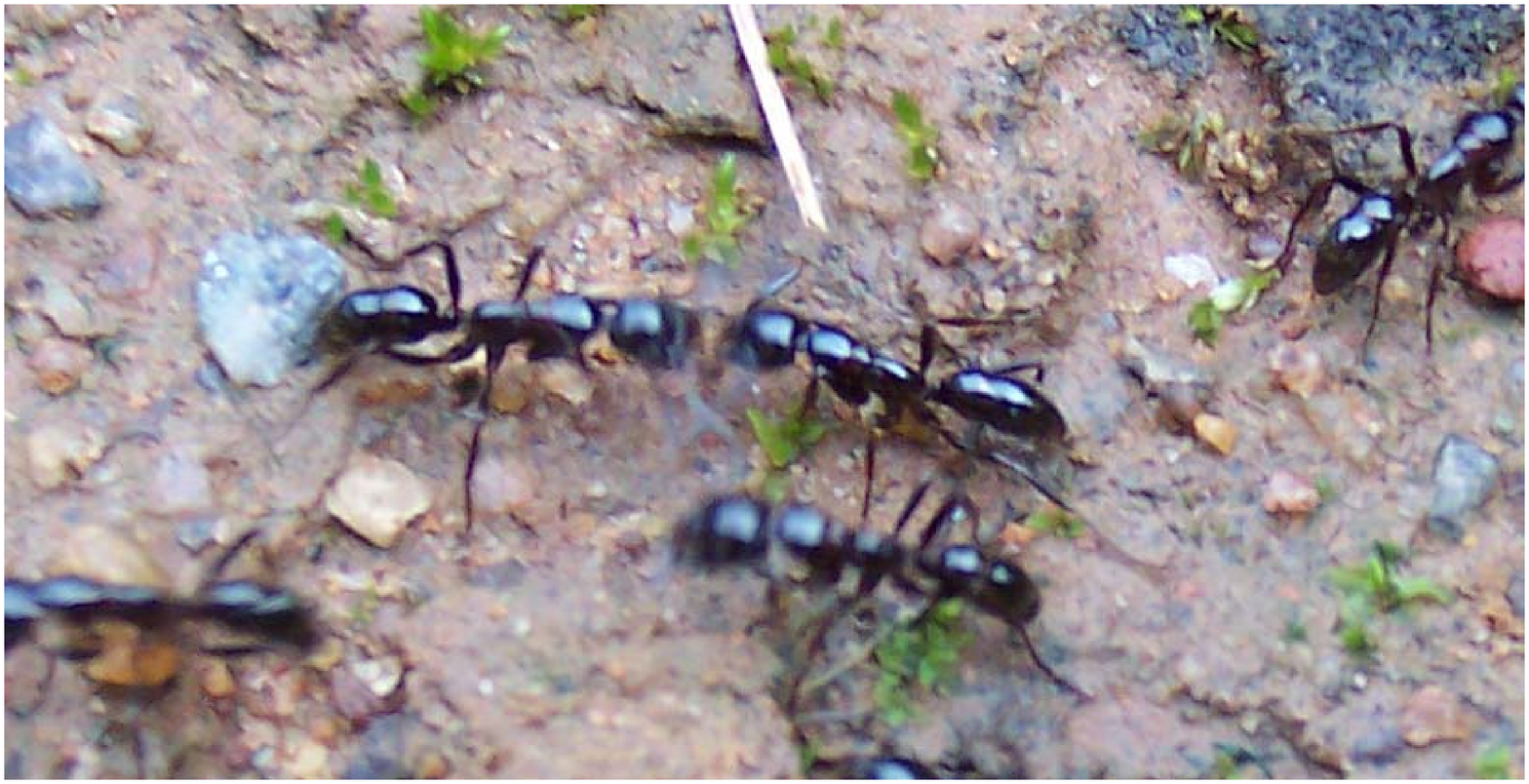}
\end{center}
\caption[]{Illustration of the observed section. Ants following 
  the established trail leave the observed section in the same order
  of entering (left). 
  Therefore the $n$-th ant entering will also be the
  $n$-th one leaving. Counting is done separately for each direction.
  Ants facing each other in opposite directions slow down in
  so-called head-on encounters (right). }
\label{eps7}
\end{figure}

\subsubsection{Basic quantities and observations}

For collecting quantitative data we chose a relatively small section
located in the middle of the trail (Fig.~\ref{eps7b}) to reduce
additional effects arising from the boundaries. For the same reason
obstacles affecting flow are avoided. The trail itself connects the
nesting site on the right to a lawn on the left.  Regarding the
ecological context, ants were not observed to carry any kind of load
like nesting material or pieces of prey. Nevertheless the colony was
still present for a few more days after data collection.

Probably the most surprising feature is the apparent absence
of overtaking. Although ants temporarily left the trail and were
passed by succeeding ones, we never observed an ant speeding up in
order to overtake. Making use of this observation one is able to
extract traffic data based on so-called cumulative counting
\cite{may}. Therefore, we use video recordings of the observed section
(Fig.~\ref{eps7}).  Counting itself had to be carried out by hand as
various video tracking systems failed. The time of entering and
leaving of every single ant is measured. Each ant produces a datapoint
for entering $(t^{j}_{+},n^{j}_{+})$ and leaving $(t^{j}_{-},n^{j}_{-})$ 
the observed section depending on the direction of movement 
$j\in\left\lbrace L,R \right\rbrace$. 
The $n$-th ant enters the section at time $t^{j}_{+}$ and leaves it 
at time $t^{j}_{-}$. $n_+(t)$ ($n_-(t)$) denotes the number of ants
that have entered (left) the measurement section up to time $t$.
The curves defined by the datapoints $(t^{j}_{+},n^{j}_{+})$ and 
$(t^{j}_{-},n^{j}_{-})$ are called arrival and departure
functions respectively (Fig.~\ref{eps8}).

One directly observable quantity is the flow of ants which is given as
the slope of the arrival or departure function (Fig.~\ref{eps8} right
inset):
\begin{equation}
\Delta t_{\pm}^{j}(n)=t_{\pm}^{j}(n)-t_{\pm}^{j}(n-1) \quad \text{and}  
\quad f_{\pm}^{j}(n)=\dfrac{1}{t_{\pm}^{j}(n)} 
\quad (j\in\left\lbrace L,R \right\rbrace)\, .
\label{th-def}
\end{equation}
Here $\Delta t_{\pm}(n)$ denotes the time-headway introduced earlier.
Under the assumption of approximately constant velocities $v^{j}(n)$
(see (\ref{vel-def})) along the observed section the distance-headway can be calculated 
additionally:
\begin{equation}
\Delta d^{j}(n)=\left( t_{+}^{j}(n)-t_{+}^{j}(n-1) \right) v^{j}(n-1)  
\quad (j\in\left\lbrace L,R \right\rbrace)
\label{dh-def}
\end{equation}
For determining density or average velocity, the length $L$ of the
observed section is needed.  By tracing the path of one single ant and
putting marks to a transparency on the video screen one obtains $L$ in
units of the body length of the ants. This method avoids errors
arising from the observer's perspective or from biasing the trail by
direct measurements.

\subsubsection{Derived quantities}

\begin{figure}[]
\begin{center}
\includegraphics[scale=0.4]{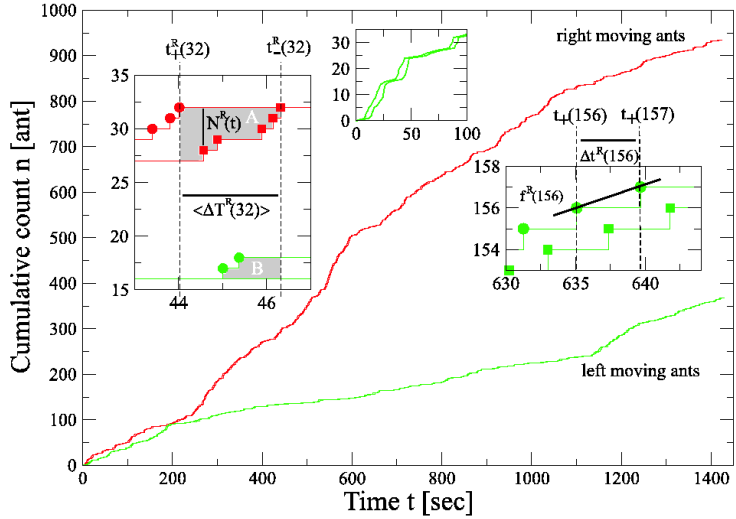}
\end{center}
\caption[]{Illustration of data extraction. Entering $n(t_{+}^{R})$ 
  and leaving $n(t_{-}^{R})$ ants produce datapoints. Based on these,
  travel time $\left\langle \Delta T^{R}(n)\right\rangle $ and average
  velocity $\left\langle v^{R}(n)\right\rangle $ are calculated. Also
  the instantaneous number of ants $N^{R}(t)$ within the observed
  section is extracted (see left inset). Based on time headway inflow
  $f^{R}_{+}(n)$ is calculated (see right inset).  Generally,
  datapoints for entering and leaving ants exhibit a pairwise
  structure (see middle inset). }
\label{eps8}
\end{figure}

As ants do not overtake, the $n$-th ant entering is also the $n$-th
one leaving (Fig.~\ref{eps7} left). Therefore, the average velocity of
the $n$-th ant while passing the observed section of length $L$ is
given by (Fig.~\ref{eps8} left inset):
\begin{equation}
\left\langle \Delta T^{R}(n)\right\rangle =t^{R}_{+}(n)-t^{R}_{-}(n) 
\quad;\quad \left\langle v^{R}(n)\right\rangle 
=\frac{L}{\left\langle \Delta T^{R}(n)\right\rangle }\,.
\label{vel-def}
\end{equation}
This method in particular makes use of the average velocity rather
than of instantaneous measurements. So the measurement is less
affected by fluctuations directly observable in the walking speed or
direction of movement.

To extract further information about the dependence of traffic flow on
the spatial distribution of the ants it is necessary to measure
density. Although we will not discuss the fundamental diagram, the
method used here has already been employed successfully in case of
strictly unidirectional traffic \cite{john-phd}.


The instantaneous particle number for one particular direction
$j\in\left\lbrace L,R \right\rbrace$ is given by
\begin{equation}
N^{j}(t)=n_{+}^{j}(t)-n_{-}^{j}(t)=const. \quad  
\text{while} \quad t\in[t_{k}^{j},t_{k+1}^{j}[\,;
\quad (j\in\left\lbrace L,R \right\rbrace)\, .
\end{equation}
Here $t^{j}_{k}$ denotes the time of an entering or leaving event for
one particular direction $j$. In both cases the number of ants is
changed by one unit. Based on this, one obtains the number of ants
$N^{j}(t)$ for each direction at a particular instant of time.  Since
this number is usually not constant during the time a specific ant
needs to pass through the observation section, some averaging over the
instantaneous particle number $N^{j}(t)$ becomes necessary.

Due to the coupling to counterflow also the number
of counterflowing ants has to be taken into account. Generally, the
total time-evolution of the trail occupation is given by a list $M$ of
entering or leaving events for both directions:
\begin{equation}
M=\left\lbrace t_{i} \right\rbrace \quad \text{with} \quad 
t_{i}\in\left\lbrace t_{k}^{R},t_{k}^{L}  \right\rbrace \, .
\end{equation}
The elements of $M$ are sorted by time. The time-average of the
instantaneous particle numbers affecting the $n$-th ant while passing
the observed section (Fig.~\ref{eps8} left inset) is given by
\begin{eqnarray}
  \left\langle N^{R}(n)\right\rangle & =& \frac{A}{\left\langle \Delta
      T^{R}(n)\right\rangle } \quad\text{with}\quad
  A=\sum_{t_{i}=t_{+}^{R}(n)}^{t_{i}<t_{-}^{R}(n)}
  N^{R}(t_{i})(t_{i+1}-t_{i})\, ,\\ 
\left\langle  N_{cf}^{R}(n)\right\rangle &=& \frac{B}{\left\langle \Delta
      T^{R}(n)\right\rangle } \quad\text{with}\quad
  B=\sum_{t_{i}=t_{+}^{R}(n)}^{t_{i}<t_{-}^{R}(n)}
  N^{L}(t_{i})(t_{i+1}-t_{i})\, .
\label{avN}
\end{eqnarray}
Here the average particle number affecting a right moving ant is
calculated and therefore the left moving ants are treated as
counterflow (Fig.~\ref{eps8} left inset). The instantaneous particle
numbers for each direction are averaged from the time of entering
$t_{+}^{R}(n)$ to the time of leaving $t_{-}^{R}(n)$ of the $n$-th right
moving ant. To facilitate comparison with the proposed models, 
we introduce dimensionless densities by
\begin{eqnarray}
\rho^{R}(n)=\frac{L}{\left\langle N^{R}(n)\right\rangle}
\quad\text{and}\quad
\rho^{R}_{cf}(n)=\frac{L}{\left\langle N^{R}_{cf}(n) \right\rangle}\, .
\end{eqnarray}
In case of $\rho^{cf}(n)<1$ the travel time $\left\langle \Delta
  T^{R}(n)\right\rangle $ is assigned to the unidirectional case and
to the bidirectional case otherwise.

\subsection{Empirical results}
\label{experiment}
We already made use of some of the qualitative results. Those are also
reflected in the quantitative measurements validating the employed
methods. The first one is the apparent absence of overtaking.
Therefore one finds a pairwise structure of datapoints of entering and
leaving ants (see Fig.~\ref{eps8} middle inset). Overall, the
departure function is roughly equivalent to the arrival function
shifted by the average travel time. Obviously, the time headways are
not changed much while passing the observed section. This indicates
stable distance-headways (see \ref{dh-def}) which will also be subject
of subsequent investigations.  In case of strictly unidirectional
traffic this feature is a consequence of ants moving in platoons, and
can easily be observed directly \cite{john-phd}.

\subsubsection{Measuring flow}

The dominance of flow of right moving ants can be observed directly on
the trail and is also indicated by the slope of the corresponding
arrival and departure functions (see Fig.~\ref{eps8}). For
$t\in[0,225]$~s flow in both directions is nearly the same. Starting
from $t\approx 225$~s the flow of right moving ants increases whereas
the flow of left moving ants decreases. Starting from $t\approx
1100$~s, flows are again nearly the same. One behavioral pattern that
affects data-collection directly are so-called U-turns. The average
rate for right moving ants is about 4$\%$ for the first two
time-intervals and reaches 7$\%$ for the last one. Left moving ants
performed U-turns constantly at a rate of about 7$\%$. As flow in both
directions is quite fluctuating this might indicate that U-turns are
performed independently of flow itself. The existence of U-turns has
also consequences for the data analysis since the pairwise structure
of datapoints required by the introduced method of measuring average
single-ant velocities has to be restored.
\begin{figure}[ht]
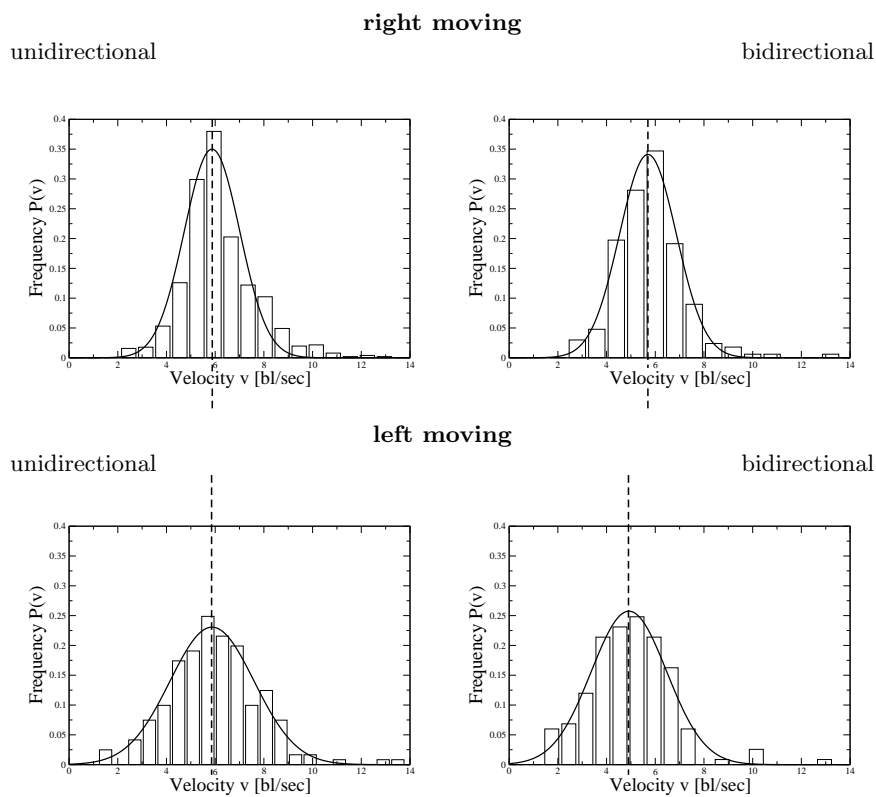

\begin{center}
\textbf{right moving}\\
unidirectional \hfill bidirectional\\
\vglue20pt
\includegraphics[scale=0.21]{LR-uni.eps}\hglue20pt
\includegraphics[scale=0.21]{LR-bi.eps}\\ \vglue1pt
\textbf{left moving}\\
unidirectional \hfill bidirectional\\
\includegraphics[scale=0.21]{RL-uni.eps}\hglue20pt
\includegraphics[scale=0.21]{RL-bi.eps}
\end{center}
\caption[]{Velocity distributions for a real bidirectional single lane 
  trail. The velocity of the right moving ants is hardly affected by
  counterflow. Average velocity as well as variance are nearly the
  same in both cases. Additionally, the mean value of left moving ants
  in absence of counterflow is nearly the same as for right moving
  ants. But in presence of counterflow a significant decrease is
  found. Generally one also observes a larger variance for left moving
  ants.}
\label{eps9}
\end{figure}

\subsubsection{Velocity- and distance-headway distributions}

For the time-interval of data collection we extracted average
velocities according to (\ref{vel-def}).  In order to investigate the
impact of counterflow we employ velocity distributions rather than
fundamental diagrams. Although those can also be calculated using the
described setup, a very large number of datapoints in comparison to
the unidirectional case is needed. One would have to cover the whole
$\rho_{L}\times\rho_{R}$-plane instead of the one-dimensional
$\rho$-line.  As a first observation, one finds that the velocity
distributions of the right moving ants exhibits a much lower variance
than for the left moving ones (see Fig.~\ref{eps7} right). Also the average
velocity shows no significant decrease in case of counterflow. This is
quite different for the left moving ants. Obviously, the average
velocity is significantly decreased by counterflow due to the right
moving ants. Overall, one finds two velocities: the reduced one for
left moving ants due to counterflow and the second one attained by all
right moving ants and for the unidirectional case of the left moving
ones.  This can be explained by making use of the advantage of
employing video recordings for data extraction. Head-on encounters are
observed for both directions (see Fig.~\ref{eps7} right). But
additionally the right moving ants take advantage of some kind of
follow-the-leader behavior. They are occupying the center of the
trail. As a consequence, left moving ants move off-center also in
absence of counterflow.  Therefore the left moving ants are more
affected by counterflow.  But also generally variance is much larger
for left moving ants. This might be caused by the lower density in
this direction as a density dependence of the velocity distribution
has been found for strictly unidirectional trails \cite{john-phd}.
From the videos one observes that the movement of the left moving ants
is slightly less directed than for the right moving ants. Therefore a
more fluctuating movement is observed. This could be a consequence of
the direct connection between trail pheromones and mass orientation
known from many other species \cite{ratnieks,ant-large}.
\begin{figure}[ht]
\begin{center}
\textbf{right moving}\\
unidirectional \hfill bidirectional\\
\includegraphics[scale=0.21]{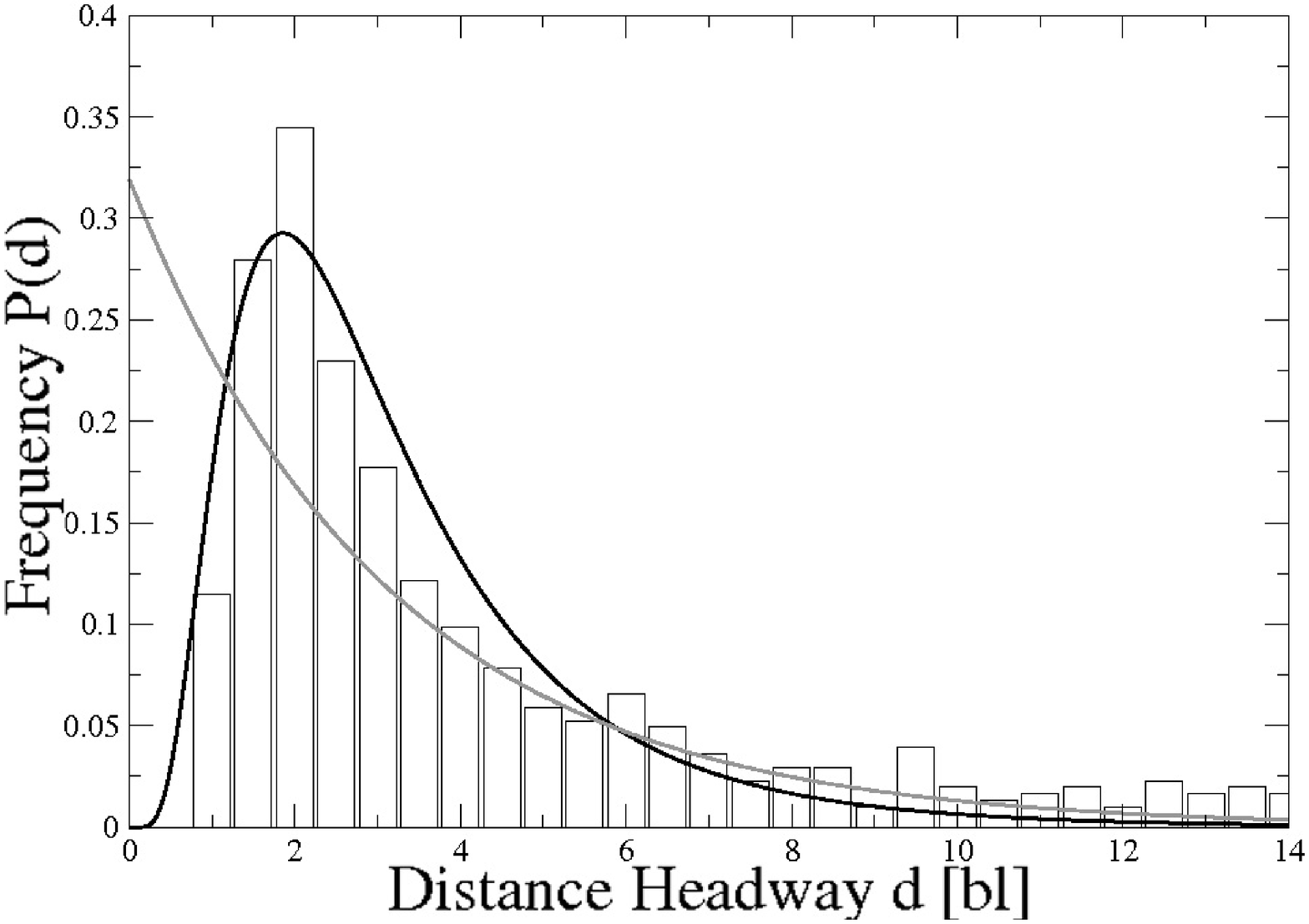}\hglue20pt
\includegraphics[scale=0.21]{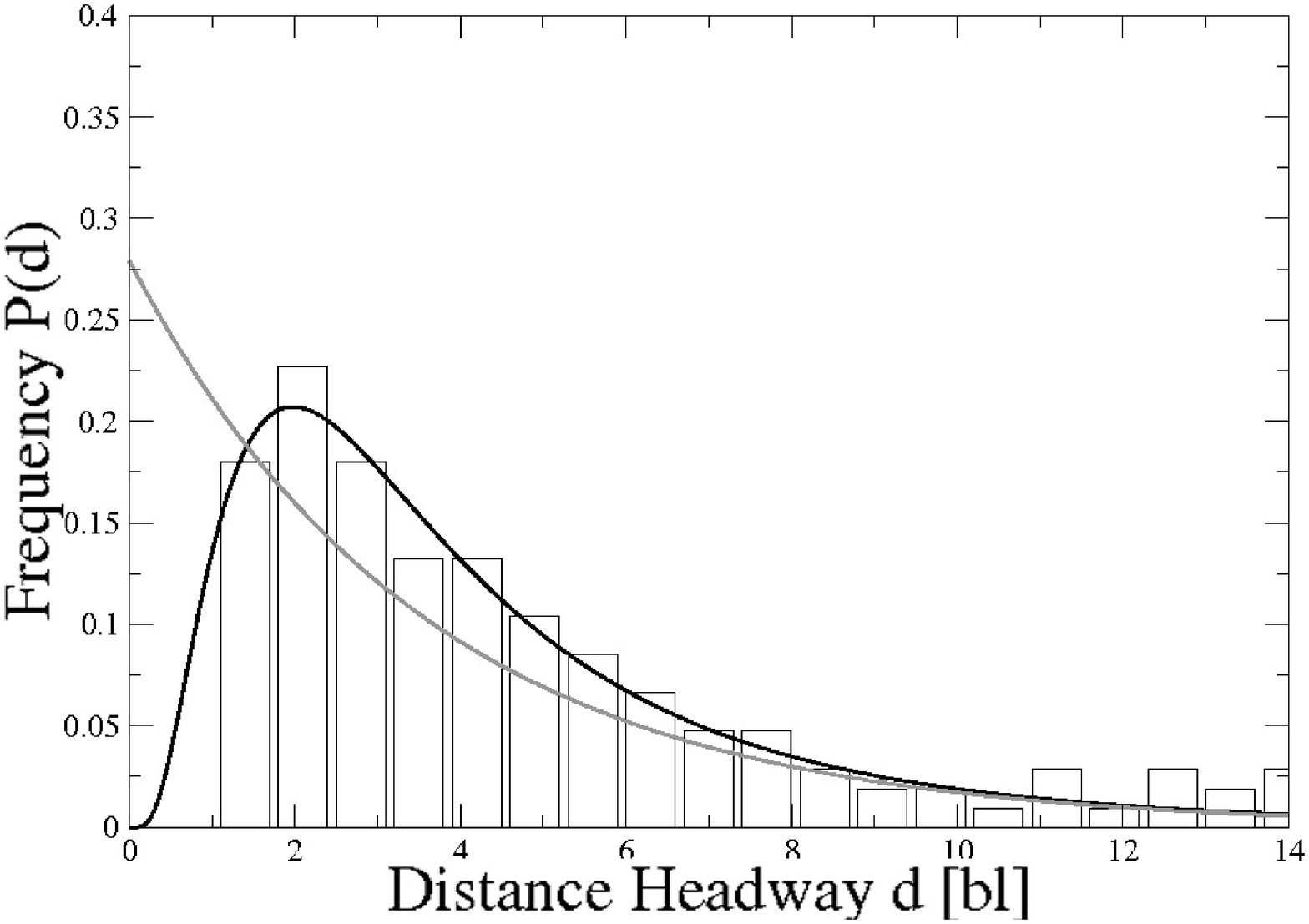}\\ \vglue1pt
\textbf{left moving}\\
unidirectional \hfill bidirectional\\
\includegraphics[scale=0.21]{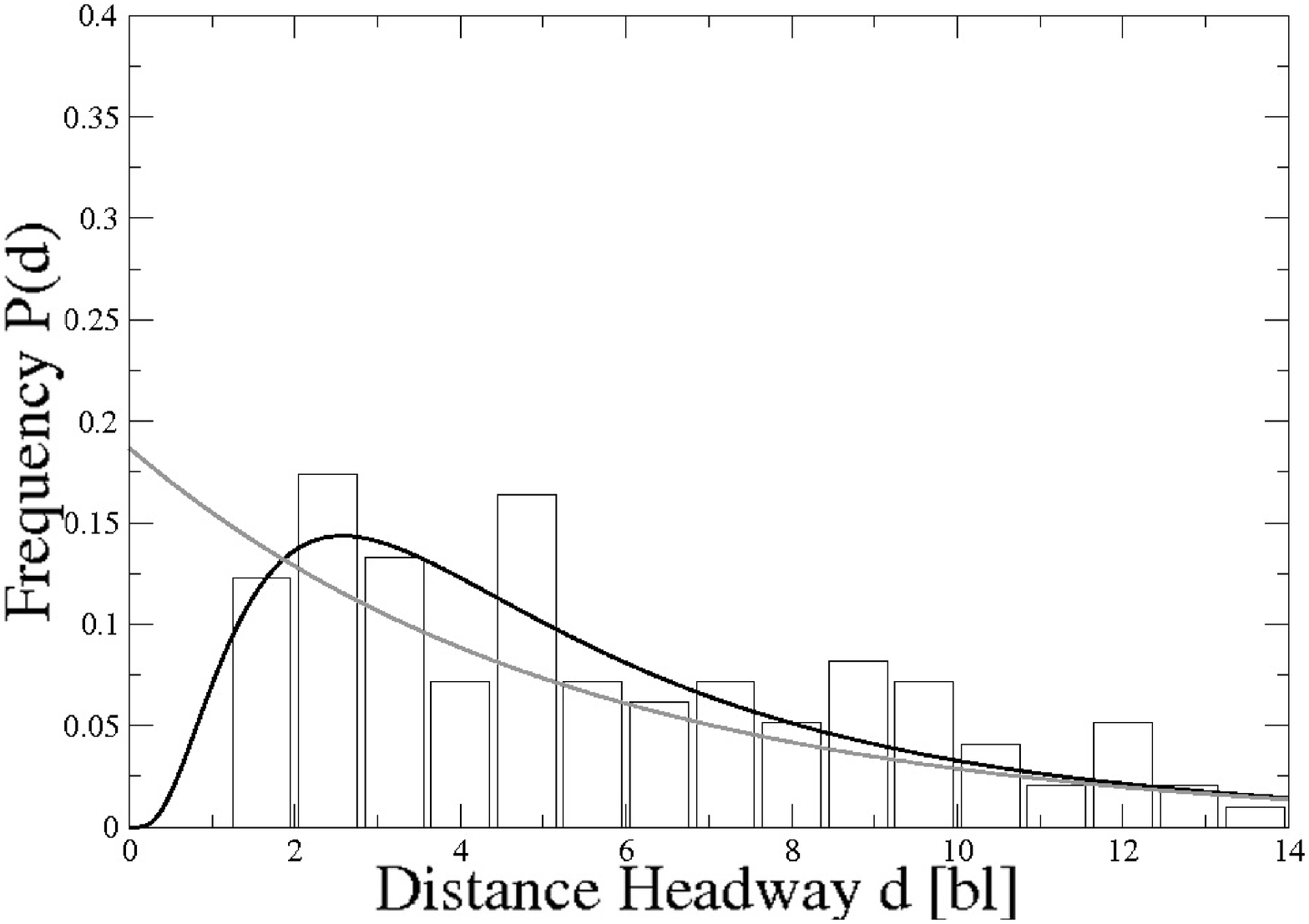}\hglue20pt
\includegraphics[scale=0.21]{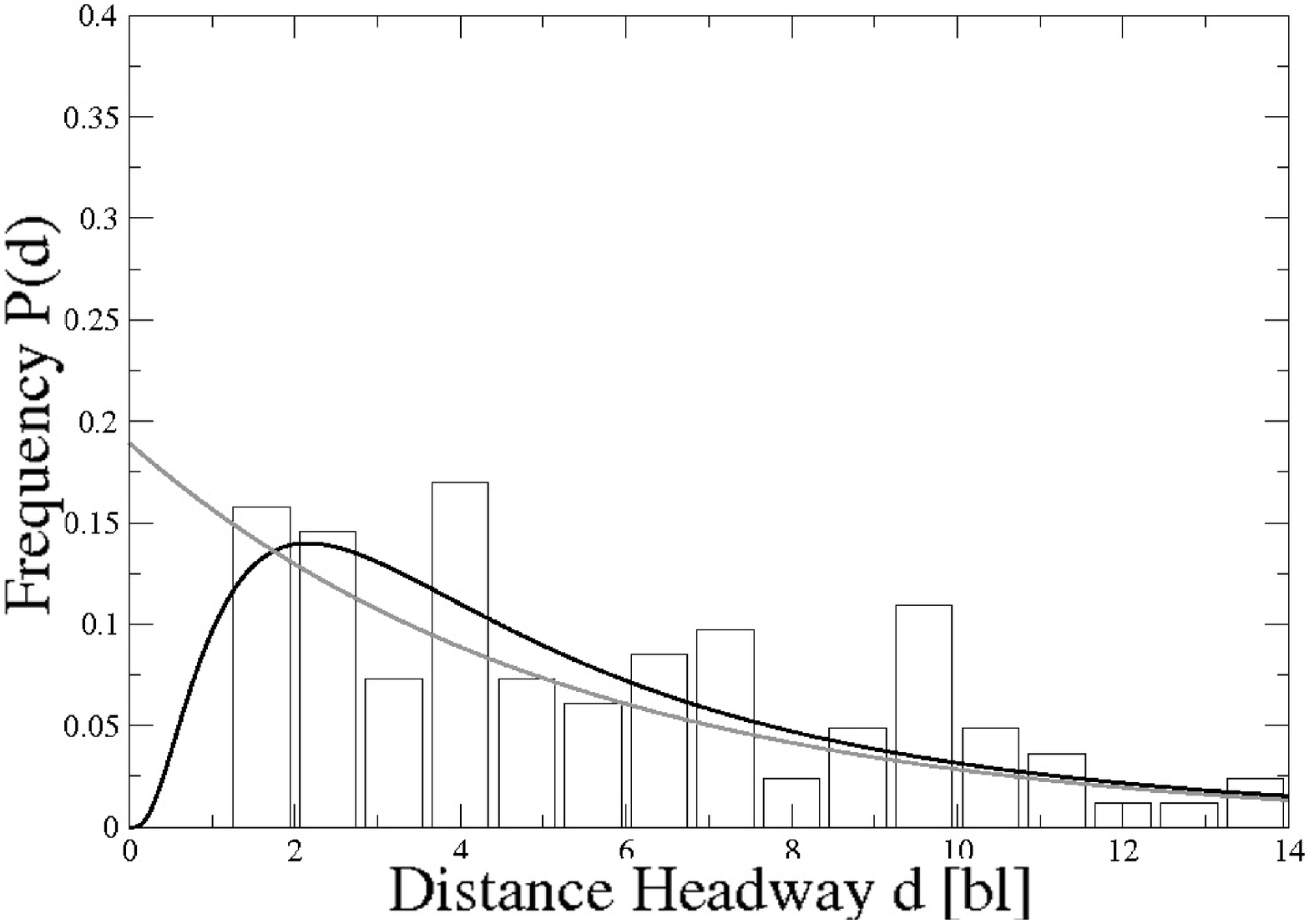}
\end{center}
\caption[]{Distance-headway distributions for a real bidirectional single lane
  trail. The left moving ants are distributed homogeneously along the
  trail. So distance-headways exhibit a negative-exponential
  distribution (grey). The right moving ants show a different spatial
  pattern. Obviously an asymmetric distribution of distance-headways
  (e.g. log-normal) is more appropriate (black).  }
\label{eps9b}
\end{figure}

Also the directly observable spatial pattern are affected as indicated
by the distribution of distance-headways (see Fig.~\ref{eps9b}).
Distances between left moving ants are randomly distributed exhibiting
a so-called negative exponential distribution.  In traffic engineering
(\cite{may}) this corresponds to a homogeneous spatial distribution of
cars on the road. With respect to our models we therefore find the
equivalent of the TASEP-regime (see Fig.~\ref{eps3} left). For the
right moving ants generally an asymmetric distribution of
distance-headways appears to be more appropriate. This indicates a
deviation form the homogeneous spatial distribution found for the left
moving ants.

\section{Summary and discussion}

We have introduced two minimal cellular automaton models for traffic
on uni- and bi-directional ant trails. Similarities and differences 
between these models and the models for vehicular traffic or pedestrians 
dynamics have been emphasized.
Especially for pedestrian dynamics there are several similarities
which also are supported by recent experiments (see e.g.\ \cite{Encycl08}
and references therein).
The main interaction in both ant-traffic models is based on different
kinds of dynamically induced disorder. In the unidirectional case
disorder is assigned to the moving agents by a virtual 
pheromone field. Although this is quite speculative (as, to our knowledge, no 
empirical evidence for such an interaction has been found yet), the main
feature of the model is quite robust. Probably most other mechanisms, 
which assign particle-wise disorder with respect to the hopping rates, 
would produce nearly the same pattern. For the bidirectional model
dynamically induced effective site-wise disorder is the key mechanism.
This mechanism is independent of the pheromone field as ants moving
in counterdirection induce a change in the hopping rates (from $q$ or
$Q$ to $K$). So this mechanism is based on the directly observable
behavior exemplified in Sec.~\ref{experiment}.

The stationary state in both models is characterized by the
spatio-temporal distribution of ants on the trail. We have shown that
the flow properties of the system are directly linked to empirically
observable aggregation patterns.  In particular, the bidirectional
model shows a rich variety of patterns. Based on the general properties
of the bidirectional model, we have also pointed out qualitative features
of additional quantities which might be of interest in case of
bidirectional flow and could also be tested experimentally.


In addition to the investigation of theoretical models, we have 
presented empirical data obtained by a simple experimental setup. 
First observations for one particular
species \textit{Leptogenys processionalis} at least qualitatively confirm 
the spatio-temporal patterns predicted by the models \cite{john-phd}
(see Fig.~\ref{eps6a}). In general, clustering phenomena
seem to be a common feature in ant colonies \cite{ant-large}. On a
quantitative level, velocity distributions and fundamental diagrams
have been obtained \cite{burd2,john-phd}. Also here the main features
of the unidirectional model have been found. For the full
bidirectional case only few experimental data are available until now
\cite{johnson,burd2}.  Although, in principle, the employed setup is
capable of obtaining the full bidirectional fundamental diagram, more
data are needed. Therefore, we have restricted the discussion to velocity-
and distance-headway distributions for investigating the impact of counterflow.
It turns out that in the case of strongly asymmetric flow, as observed here, 
ants move differently depending on the direction. With respect to our models
we would need to introduce different hopping rates for each direction.
Nevertheless, the mutual slowing down, as predicted by the
models, could be confirmed. Also the impact of counterflow on the spatial 
distribution of ants along the trail was shown. 


The empirically as well as the theoretically observed organization of
traffic flow raises fundamental questions on the advantage of such
organizations. Both cellular automata models produce moving clusters
for an appropriate choice of parameters.  In the unidirectional model
this happens for a relatively wide density regime.  The existence of
these clusters corresponds to the regime of constant average velocity.
In the TASEP the average velocity decreases strictly monotonically
with increasing density as ants are distributed homogeneously along
the trail. So moving in clusters enables the ants to keep on moving at
a higher velocity in comparison with a homogeneous distribution in
TASEP. This is obviously achieved by reducing mutual blocking. One
might argue that the fundamental diagram shows a maximum value of
average velocity. This maximum is attained at the point of sharp
increase from the cluster- to the homogeneous distribution.
Nevertheless, a maximum of the average velocity at this point would be
quite unstable.  Even small fluctuations in density would lead to
large fluctuations in the average velocity. In comparison, the cluster
regime is quite stable against fluctuations in density.  Therefore, in
natural systems clustering is expected to lead to a decrease of travel
time which can be seen as a user optimum.

In the bidirectional case, moving clusters only occur at very low
densities. The underlying mechanism is still the same as in the
unidirectional model.  At intermediate to high densities the main
feature, namely the formation of localized clusters leading to a
constant value of flow, emerges. This feature solely depends on the
presence of counterflow.  In this situation flow is the crucial
quantity. In an appropriate ecological context, the outbound flow for
example determines the number of ants travelling to the food source,
whereas inbound flow is related to the amount of food carried back to
the nest. Very different values of inbound and outbound flow would
lead to a too large or too small number of ants at the source. So a
vanishing effective flow, which is roughly independent of density
fluctuations, also ensures a constant number of workers at a particular
destination. As this feature involves flow, it can be interpreted as
some kind of system optimum.\\ 

\parindent0pt
{\bf Acknowledgments}\\

The authors would like to thank M. Burd, P. Chakroborty, R. Gadagkar,
B. H\"olldobler, A. Kunwar and T. Varghese for informative
discussions. They also thank the referees and the editor for
helpful suggestions concerning the presentation.
Parts of the presented work of one of the authors (AJ)
has been supported by the German Academ
through a joint Indo-German research project.


\end{document}